\documentclass{emulateapj}

\shorttitle{ XLF GALAXIES} 
\shortauthors{Norman et al.} 
\input psfig.sty 
\slugcomment{Accepted for publication in ApJ} 
 
\begin{document} 
 
\title{THE X-RAY DERIVED COSMOLOGICAL STAR FORMATION HISTORY AND THE 
GALAXY X-RAY LUMINOSITY FUNCTIONS IN THE {\bf \it Chandra} DEEP FIELDS 
NORTH AND SOUTH} 
  
\author {Colin Norman\altaffilmark{1,2,3}, Andrew 
  Ptak\altaffilmark{1}, Ann 
  Hornschemeier\altaffilmark{1,4}, Guenther Hasinger\altaffilmark{5},  
  Jacqueline Bergeron\altaffilmark{6}, 
  Andrea Comastri\altaffilmark{7},  
  Riccardo Giacconi\altaffilmark{1,8}, Roberto 
  Gilli\altaffilmark{9}, Karl Glazebrook\altaffilmark{1}, Tim 
  Heckman\altaffilmark{1},  
  Lisa Kewley\altaffilmark{10},Piero Ranalli\altaffilmark{11}, Piero 
  Rosati\altaffilmark{1,3}, Gyula Szokoly\altaffilmark{5}, Paolo 
  Tozzi\altaffilmark{12}, JunXian Wang\altaffilmark{1}, Wei 
  Zheng\altaffilmark{1}, Andrew Zirm\altaffilmark{13}} 
 
\altaffiltext{1}{The Johns Hopkins University, Homewood Campus, 
  Baltimore, MD 21218} 
 
\altaffiltext{2}{Space Telescope Science Institute, 3700 San Martin Drive, Baltimore, MD 21218} 
 
\altaffiltext{3}{European Southern Observatory, Karl-Schwarzschild-Strasse 2, Garching, D-85748, Germany} 
 
\altaffiltext{4}{Chandra Fellow} 
 
\altaffiltext{5} {Max-Planck Institute for extraterrestrische Physik, Giessenstrasse, 85740 Garching bei Munich, Germany} 
 
\altaffiltext{6}{Institut d'Astrophysique, Bd. Arago, bis 98, Paris, 75014, France} 
 
\altaffiltext{7}{INAF-Osservatorio Astronomico di Bologna,
via Ranzani 1, I--40127 Bologna, Italy} 
 
\altaffiltext{8}{Associated Universities, Inc. 1400 16th Street, NW, Suite 730, Washington, DC 20036} 
 
\altaffiltext{9}{Osservatorio Astrofisico di Arcetri, Largo E. Fermi 
5, 50125 Firenze, Italy} 
 
\altaffiltext{10}{Harvard-Smithsonian Center of Astrophysics, MS-20, 
  60 Garden Street, Cambridge, MA 02138} 
 
\altaffiltext{11}{Universit\`a di Bologna, Dipartimento di Astronomia, 
via Ranzani 1, I--40127 Bologna, Italy} 

\altaffiltext{12}{Osservatorio Astronomico, Via G. Tiepolo 11, 34131 Trieste, Italy} 
 
\altaffiltext{13}{Leiden Observatory, P.O.Box 9513, NL-2300 RA, Leiden, The Netherlands} 
 
\begin{abstract} 
 
The cosmological star formation rate in the combined Chandra Deep 
Fields North and South is derived from our X-Ray Luminosity Function 
for Galaxies in these Deep Fields. Mild evolution is seen up to 
redshift order unity with SFR $\sim (1 + z)^{2.7}$. This is 
the first directly observed {\it normal 
star-forming galaxy} X-ray luminosity function (XLF) at 
cosmologically interesting redshifts (z$>$0).  This provides the most 
direct measure yet of the X-ray derived cosmic star-formation history 
of the Universe.

We make use of Bayesian statistical methods to classify the galaxies 
and the two types of AGN, finding the most useful discriminators to be 
the X-ray luminosity, X-ray hardness ratio, and X-ray to optical flux 
ratio. There is some residual  
AGN contamination in the sample at the bright end of the luminosity 
function. Incompleteness slightly flattens the XLF at the faint end of 
the luminosity function. 
\\ 

The XLF has a lognormal distribution and agrees well with the radio 
and infrared luminosity functions. However, the XLF does not agree 
with the Schechter luminosity function for the H$\alpha$LF indicating 
that, as discussed in the text, additional and different physical 
processes may be involved in the establishment of the lognormal form 
of the XLF. 
\\

The agreement of our star formation history points with the other star
formation determinations in different wavebands (IR, Radio, H$\alpha$)
gives an interesting constraint on the IMF.  The X-ray emission in the
Chandra band is most likely due to binary stars although X-ray
emission from non-stellar sources (e.g., intermediate-mass black holes
and/or low-luminosity AGN) remain a possibility.  Under the assumption
it is binary stars the overall consistency and correlations between
single star effects and binary star effects indicate that not only is
the one parameter IMF(M) constant but also the bivariate IMF($M_1,
M_2$) must be constant at least at the high mass end.  Another way to
put this, quite simply, is that X-ray observations may be measuring
directly the {\bf \it binary star formation history} of the Universe.
\\ 
 
X-ray studies will continue to be useful for probing the star 
formation history of the universe by avoiding problems of 
obscuration. Star formation may therefore be measured in more detail 
by deep surveys with future x-ray missions. 
 
\end{abstract} 
 
 
\section{Introduction} 
   
There have been many recent studies of star formation in galaxies and
of the star formation history of the universe derived from data in the
radio, IR, and optical \citep{lilly1996, madau1998, rowanrobinson1997,
haarsma2000, Cole01,Baldry02,teplitz2003}. In the range from the
present epoch to redshifts of order unity, recent critical
compilations and discussions of \citet{sullivan2001},
\citet{hopkins2003}, \citet{sullivan2004} and \citet{hogg2004} show
that the results from the multi-waveband data have a dispersion of 1-2
orders of magnitude in the comoving cosmic star formation density . As
noted by these authors, there are important physical corrections that
need to be made to go from the observations in a particular band to
the cosmic star formation rate which include physical understanding of
the dust extinction, the Initial Mass function (IMF), and stellar
population models. Reasonable interpretations of the current
observations in this redshift range have been presented that argue, on
the one hand, that there is a steep decline in the star formation rate
to the present epoch \cite{hogg2004} or, on the other, that the cosmic
star formation density has a shallow decline in the same redshift
range \cite{wilson2002}. Therefore it is important to utilize all
wavebands to study this phenomenon from different aspects and with
different selection effects. The X-ray band has now just opened up to
detailed studies of star formation in galaxies at cosmological
distances.
 
Hitherto, even deep X-ray surveys studied only the cosmological 
populations of evolving active galaxies and quasars. X-ray studies of 
individual nearby galaxies were performed and the underlying hot gas 
and stellar x-ray source components analyzed \citep{fabbiano1989}. 
However, the deep 1-2 Megasecond surveys in the Chandra Deep  
Field South and North, respectively, now show a major cosmological   
population (in the range from present day to redshifts of order unity) 
of X-ray emitting normal star forming galaxies at faint flux 
levels. Stacking analysis allows one to extend the x-ray properties 
studied to even larger redshift ranges \citep{hornschemeier2002}.  
 
Some notable results on star-forming galaxies have already been
derived in deep X-ray survey work.  In the $Chandra$ Deep Field-North
(hereafter CDF-N) it has been found that the faint 15$\mu$m
population, composed primarily of luminous infrared starburst galaxies
(e.g. \citet{chary2001}) are associated with X-ray-detected
emission-line galaxies (Alexander et al. 2002).  {\it Chandra} and
{\it XMM-Newton} stacking analyses of relatively quiescent
(non-starburst) galaxies have constrained the evolution of X-ray
emission with respect to optical emission to at most a factor of 2--3
(e.g., Hornschemeier et al. 2002; Georgakakis et al. 2003).  Studies
of the quiescent population of galaxies has also provided some initial
constraints on the evolution of star-formation in the Universe ( SFR
$\sim (1+z)^k$ for $z <<1$ where k $<3$; Georgakakis et al. 2003).
 
We now 
discuss some relevant details of the X-ray observations. 
We derive X-ray Luminosity Functions and the corresponding 
 cosmological star formation history using X-ray data from the 2~Ms
 and 1~Ms exposures of the CDF-N and {\it Chandra} Deep Field South
 (hereafter CDF-S).  This analysis requires the extreme depth of the
 CDF surveys as non-active galaxies arise in appreciable numbers at
 extremely faint X-ray fluxes; the fraction of X-ray sources that are
 normal and star-forming galaxies rises sharply below 0.5--2~keV
 fluxes of $\approx1\times10^{-15}$~erg~cm$^{-2}$~s$^{-1}$
 \citep[e.g., Figure~6 of][; see also the fluctuation analysis results
 of Miyaji \& Griffiths 2002]{hornschemeier2003}.
 
The luminosity functions for galaxies in the radio, optical and
infrared have been studied extensively (c.f. \citet{tresse2002} for a
recent discussion). Prior to this work, the X-ray luminosity function
for normal galaxies was estimated from the optical galaxy luminosity
function by \citet{georgantopoulos1999}.  Schmidt, Boller \& Voges
(1996) reported on a galaxy XLF derived in the local ($v <
500$~km~s$^{-1}$) Universe as well, however their sample does include
some fairly active galaxies (e.g., Cen A) and does not appear to be an
entirely clean normal/star-forming galaxy luminosity function.  Our
work seeks to construct a relatively clean {\it normal star-forming }
galaxy luminosity function, and possesses the great advantage of doing
this at cosmologically interesting distances (our two luminosity
functions have median redshifts $z\approx0.3$ and $z\approx0.8$).  We
may thus study the {\it evolution} of X-ray emission from galaxies
with respect to cosmologically varying quantities such as the global
star-formation density of the Universe.  Our galaxy XLFs have their
own set of selection effects that we discuss later but they can add to
the overall picture of the physics of galaxy evolution. For example,
it is obvious that corrections for obscuration will not be as
important here as in the analysis of H$\alpha$ Luminosity functions
discussed in detail below.
 
We expect correlations between the various classic cosmological
indicators of star formation (H$\alpha$, IR, radio) and our X-ray
studies. In general, the fluxes we measure in the respective wavebands
associated with the different methods are all ultimately connected
with the evolution and death and transfiguration of massive stars.
For example, the radiative luminosity of OB stars, the mechanical
energy input from supernovae and the accretion power of high mass
X-ray binaries (HMXBs) are all ultimately derived from massive stars.
It is therefore not surprising that tight correlations have been found
empirically between the X-ray flux, the radio flux and the IR
continuum.  The most recent careful analysis in the local Universe is
from \citet{ranalli2003} which is discussed in detail
below. Interestingly, the correlations appear to hold for at least the
radio and X-ray bands at high redshift \cite{bauer2002,grimm2003}. We
use these latest empirically derived correlations betwen radio, IR,
and X-ray star formation indicators as the central relations that
allow us to infer star formation rates from the X-ray data on galaxies
in the Chandra Deep Fields North and South. In addition, it is useful
to also consider theoretical studies on the cosmic X-ray evolution of
galaxies which have been presented by \citet{cavaliere2000},
\citet{ptak2001} and \citet{ghosh2001}.
 
Observations of detailed X-ray stellar populations in individual
nearby galaxies have been carried out by
\citet{zezas2002},\citet{kilgard2002} and \citet{colbert2003}.  Using
archival $Chandra$ data on 32 galaxies in the nearby Universe, Colbert
et al. (2003) have established that the X-ray emission from accreting
binaries in galaxies is correlated with the current level of star
formation in their host galaxies.  This relation appears to be both
macroscopic in that the entire X-ray point source luminosity of
galaxies scales with star-formation rate and microscopic in that the
slope of the X-ray binary luminosity function {\it within} galaxies
correlates with star formation.
 
In our studies, much care has gone into accurately selecting the 
population of normal star forming galaxies. Contamination by AGNs is a 
potentially serious problem. We have a multi-parameter probability 
space for the selection of the normal star forming galaxies and we have 
therefore used Bayesian methods to classify the galaxies and the two 
types of AGN. The most useful discriminators are the X-ray luminosity 
and X-ray hardness ratio as well as the optical to X-ray ratios.  
AGN contamination is most serious at the bright end of the luminosity 
function. Our analysis is a step towards a fully modern 
statistical Monte Carlo Markov-Chain study but we need to have a 
better idea of the overall structure of the probability space before we 
undertake this (c.f. \citet{hobson2002} and \citet{hobson2003}). Our 
analysis assumes, as a first step, that the priors are Gaussian and we 
show that this is a reasonable and useful assumption. 
 
These studies of the cosmological evolution of normal star forming
galaxies in the x-ray bands are the first of many detailed studies
that can be undertaken with future missions such as XEUS. Here we have
made a start on the basic luminosity functions and cosmic star
formation histories.
 
The organization of this Paper is as follows: in section 2 we 
describe the data acquisition, the selection of the galaxy sample and 
the processing of the data. Section 3 presents the derivation of the 
galaxy XLF. In Section 4, we carefully compare the galaxy XLF derived 
here with the IR LF using the recent empirically derived analysis from 
\citet{ranalli2003} (hereafter RCS). A similar 
procedure is undertaken for comparison with the H$\alpha$LF.  We also 
give the derived cosmic star formation rates from the XLF.  In 
section 5 we discuss the implications of our results and the 
potential consequences for future missions. In section 6 we give our four principal conclusions. Throughout this paper, if 
not explicitly stated otherwise, we use $H_0$ = 70 km s$^{-1}$ 
Mpc$^{-1}$, $\Omega_m = 0.3$ and $\Omega_{\Lambda}=0.7$.

\section{Data Acquisition, Selection of Galaxies, and Data Reduction} 
 
\subsection{ Data Acquisition} 
 
\subsubsection{CDF-S} 
 
From October 1999 to Dec 2000, 11 individual exposures of the CDF-S
were performed with the {\it Chandra} ACIS instrument, resulting in a
1~Ms exposure.  The sensitivity of this deep exposure has reached flux
limits of $5.5\times 10^{-17}$~erg~cm$^{-2}$~s$^{-1}$ in the soft band
(0.5--2~keV) and $4.5 \times 10^{-16}$~erg~cm$^{-2}$~s$^{-1}$ in the
hard band (2--10~keV).  At these flux levels, $>$ 80$\%$ of the cosmic
X-ray background in both bands is resolved.  A total of 346 sources
has been detected \citep{rosati2002, giacconi2002}; the X-ray data
reduction, source detection, and X-ray source catalog can be found in
\citet{giacconi2002}.  The optical spectroscopic follow-up
observations were obtained using FORS1/FORS2 on the VLT for the
possible optical counterparts of 238 X-ray sources, yielding
spectroscopic redshifts for 141 X-ray sources.  The optical spectra
and the redshifts are presented in \citet{szokoly2004}.  Zheng et al.,
in preparation, have used the ten near-UV, optical, and near-infrared
bands to estimate photometric redshifts for 342 (99\%) of the CDF-S
sources, making detailed comparison with the spectroscopic redshifts.
In some cases, the optical spectroscopic redshifts of Szokoly et
al. were not highly confident and Zheng et al. was able to place a
better redshift constraint using the multiwavelength SED of the
sources.  We therefore used those redshifts of Zheng et al. which we
consider to supercede those of Szokoly et al.
 
\subsubsection{CDF-N} 
 
 From November 1999 to February 2002, 20 individual exposures of the 
 CDF-N were performed by the {\it Chandra} ACIS instrument, resulting 
 in a 2~Ms exposure. In the central parts of the CDF-N the sensitivity
 reaches $2.5 \times 10^{-17}$ erg cm$^{-2}$ s$^{-1}$ in the
 0.5--2~keV band 
 and $1.4 \times 10^{-16}$ erg cm$^{-2}$ s$^{-1}$ in the 2--8~keV 
 band.  %A total %of 370 high-significance sources were detected in 
 the 1~Ms survey; 503 sources were detected in the 2~Ms data, and the 
 X-ray data reduction, source detection, and X-ray source catalog can 
 be found in \citet{alexander2003}.  The optical spectroscopic and 
 photometric follow-up observations were obtained using Keck and 
 Subaru \citep{barger2002,barger2003}.  Spectroscopic redshifts have 
 been obtained for 284 of CDF-N X-ray sources; using the multiwavelength 
photometric data, photometric redshifts were obtained for an additional 
78 sources \cite{barger2003}.  We use these 362 redshifts in our study.  
Note that although the overall completeness is roughly 71\%, including 
the photometric redshifts, the spectroscopic completeness alone for 
R $\leq 24$  is 87\%.     The optical and 
 near-infrared photometry, spectroscopic redshifts, and 
 photometric redshifts are presented  
in \cite{barger2003}.   
  
\subsection{Galaxy Selection} 
 
In our study it is essential to classify galaxies correctly.  We have 
employed two different galaxy classification techniques.  The first 
method uses standard optical classification for sources with 
high-quality optical spectra. The second approach utilizes a 
multi-parameter Bayesian analysis. 
 
We have chosen to construct XLFs in the 0.5--2~keV band since for 
moderately obscured AGN with significant star formation present, the 
soft band is dominated by star forming processes 
\citep{ptak1999,levenson2001,terashima2002}.

\subsubsection{Optical Spectroscopic Classification} 
 
Very few galaxies have optical spectra with good enough signal-to-noise to 
allow classifications based on emission-line flux ratios.  Therefore, 
it is essential  
that we develop a classification scheme which is not dependent (or overly 
dependent) upon optical spectroscopic classification.  However, there is 
important optical spectroscopic information contained within the spectra. 
For the CDF-S spectra with sufficient signal-to-noise, we have used 
the diagnostic diagrams of \citet{rola1997} in a manner similiar to  
\citet{kewley2002}.  Where only one or two emission lines are present, 
we classify the objects as AGN if broad features or high ionization lines 
were present and galaxies otherwise.  Note that the classifications in 
most of the optical spectroscopic follow-up work in the Chandra Deep 
Fields, e.g., \citet{szokoly2004}, are based upon {\it both} X-ray 
(luminosities and hardness ratios) and optical properties.  We do not 
consider the X-ray properties in classifying these galaxies as we are 
using them to construct our X-ray prior. 
 
There are 29 galaxies whose emission-line ratios are consistent with 
starbursts and/or normal galaxies rather than AGN.  We further exluded 
two of these sources since they were found at off-axis angle in excess 
of 8' where the spectroscopic completeness considerably worsens 
relative to the center of the survey (Szokoly et al. 2004).  This is thus our 
``spectroscopic CDF-S sample" with a total of 27 galaxies.  Note that 
we do not classify an analogous  
``spectroscopic CDF-N sample" as those optical spectra are not available, 
and also are generally not flux calibrated \citep{barger2003}, which 
would prevent line ratios from being calculated. 
 
\subsubsection{ Multi-parameter Classification: $(f_x/f_R)$, HR, $L_X$}  
 
Even with optical classification there is still the risk of AGN 
contamination, particularly for type-2 AGN where the AGN may be 
obscured and/or the spectral aperture encompasses the entire galaxy 
\citep[e.g.,][]{moran2002}.  Therefore, we used a second   
selection approach based on the statistical properties of the sources. 
 
For this statistical analysis we concentrated on the CDF-S 
multiwavelength data since again we had access to the optical spectra 
of potential counterparts, and could therefore most reliably identify 
control sets of galaxies and AGN.  We investigated the distributions 
of the X-ray hardness ratios and 1.4 GHz, K-band, R-band and soft 
X-ray luminosities and inferred that a good discriminator between AGN 
and galaxies was based on the X-ray/optical luminosity ratio, the 
X-ray hardness ratio and X-ray luminosity as had been found previously 
\citep[][and references therein]{hasinger2003}. The ratios 
$L_X/L_{1.4\ \rm GH}$ and $L_X/L_{\rm K}$ also show promise and will 
be explored in future work. 
 
\begin{figure} 
%\plotone{select_25may.ps} 
%\plotone{hr_ls_Q3.ps} 
\plotone{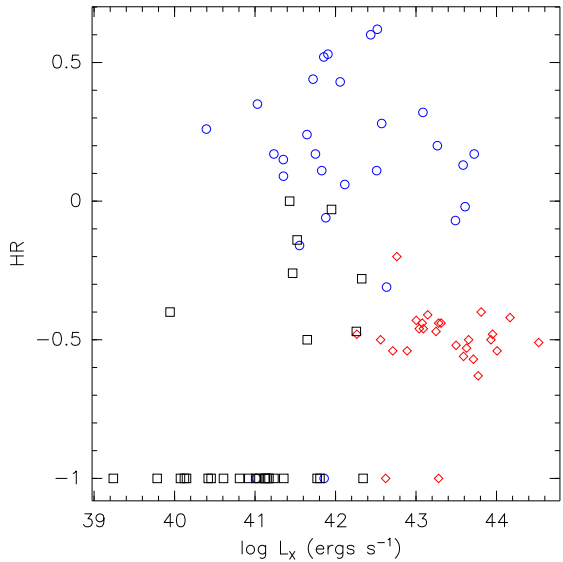} 
\figcaption{ 
X-ray hardness ratio HR vs. soft X-ray (0.5 -- 2.0 keV) luminosities 
for CDF-S sources with spectroscopic redshifts and classification.   
Galaxies are plotted as black squares, AGN1 are red diamonds and AGN2 
are blue circles. 
%Open triangles are normal galaxies with spectroscopic 
%identifications (qual $\geq 3$) and filled triangles  
%are normal galaxies with photometric identifications.  Filled circles 
%are AGN1 and filled boxes are AGN2.   
All luminosities are calculated 
assuming $H_0$ = 70 km s$^{-1}$ Mpc$^{-1}$, $\Omega_m = 0.3$ and 
 $\Omega_{\Lambda}=0.7$ . 
\label{fig:hrtype}} 
\end{figure}

In Figure \ref{fig:hrtype} we plot the hardness ratio vs soft band 
X-ray luminosity of all CDF-S sources with good spectroscopic 
redshifts and classifications.  
Different symbols are used for the 
various optical spectroscopic classifications (type 1 AGN, hereafter 
AGN1; type 2 AGN; hereafter AGN2 and normal galaxies).   
%*%*%*%*% Deleted 12/19/03 Since I don't think this is nec. 
%All sources 
%with log $L_X > 42.5$ ergs s$^{-1}$ are considered to be AGN.  
It is 
evident that the AGN1 and galaxies have similar hardness ratios and 
AGN1 have significantly higher X-ray luminosities. AGN2 span a larger 
range in X-ray luminosity than AGN1 and on average have harder spectra 
than AGN1 and galaxies. Both of these effects are expected since AGN2 
are generally X-ray-obscured sources. A brief synopsis of the 
technique follows. 

\begin{figure} 
\plotone{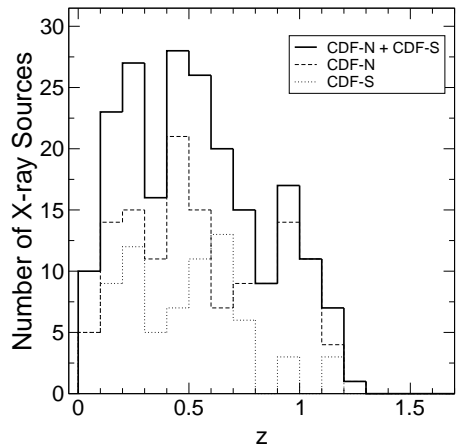} 
\figcaption{The redshift distribution of the Bayes-selected galaxy 
  sample. \label{zbayes}} 
\end{figure}

In most cases the errors on the hardness ratios are large for sources 
in the galaxy regime as there were often only a few X-ray counts 
(i.e., $\log{L_X} < 42$ and HR $< 0$) and, in fact, many sources in this 
regime only have upper-limits for HR.  Our selection approach takes 
into account the errors on the hardness ratios and the parent 
distribution properties discussed above.  This approach is described 
in Appendix A, and we also restricted our sample to sources with 
redshifts $\leq 1.2$ (since the redshift estimates for the sources 
become increasingly uncertain at higher redshifts and our two redshift 
bins should be of comparable size).  This resulted in 74 CDF-S and 136 
CDF-N sources being  
classified as galaxies. In Figure~\ref{zbayes} we show the redshift 
distribution for the (Bayesian) galaxy sample.   
 
\section{Luminosity Functions} 
 
\begin{figure*} 
\plottwo{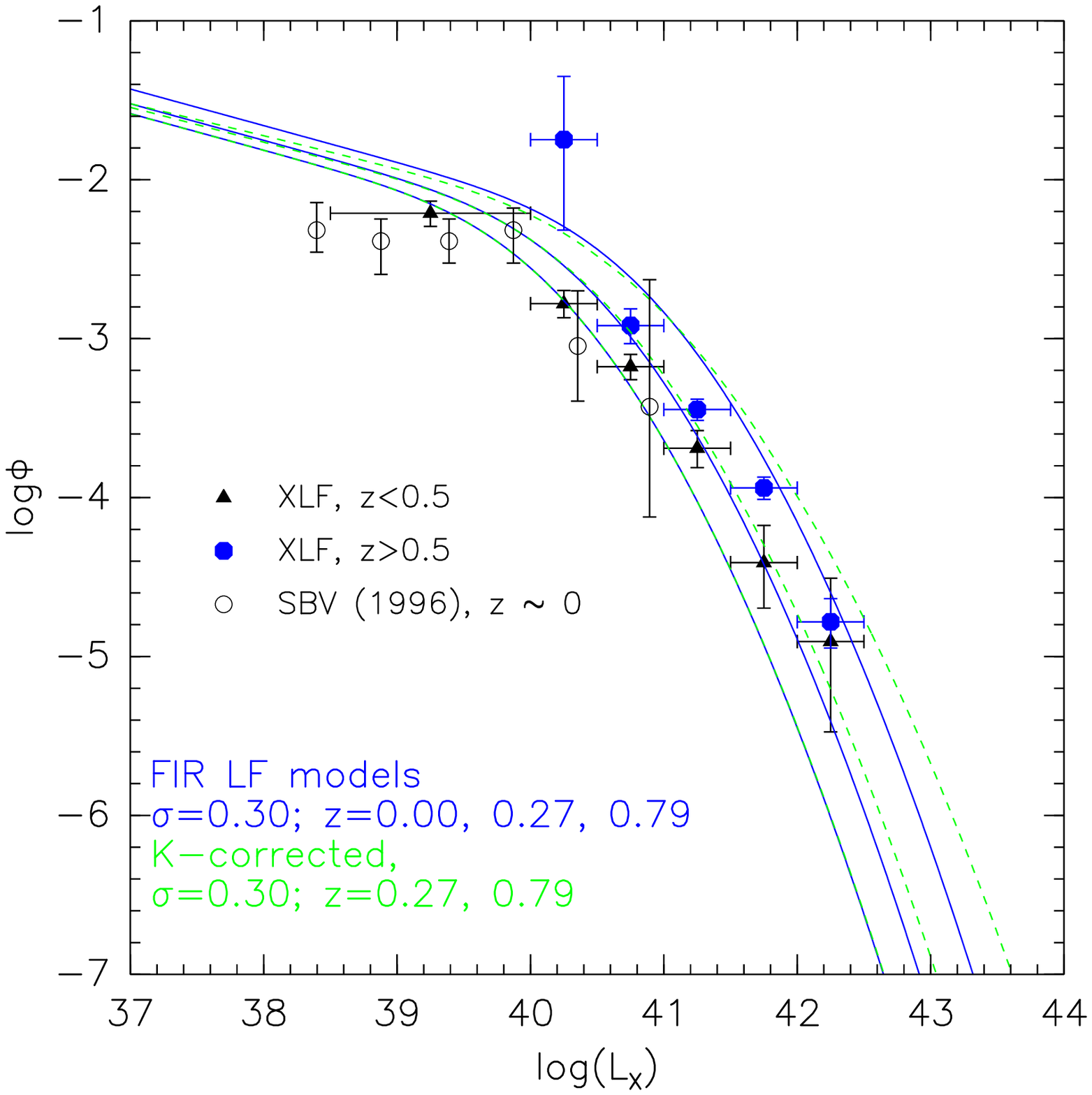}{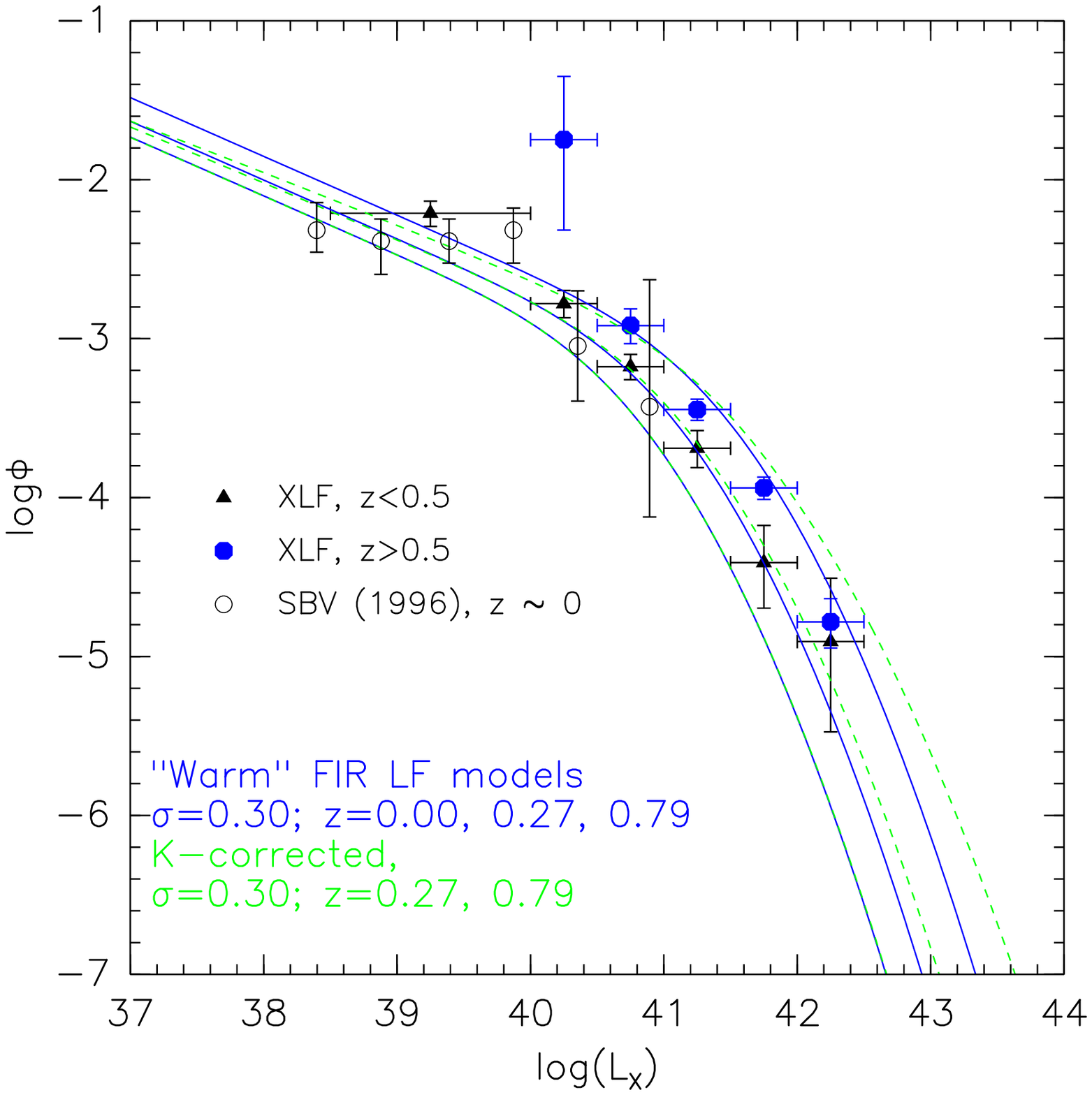} 
\figcaption{Combined CDF-N and CDF-S XLF based on the Bayesian model testing 
  sample (see text for details) along 
  with the 60$\mu$m LF from \citet{ta03} converted to the 0.5-2.0 keV 
  bandpass using the $L_{60 \mu m}$/X-ray correlation (based on the galaxy 
  sample given in Ranalli et  
  al. [2003]).  The left panel shows the 60$\mu$m LF model based on 
  the full IRAS PSCz sample while right panels shows the LF based on 
  the ``warm'' galaxy sample (with $S_{100 \mu m}/F_{60 \mu m} < 2.1$). 
  The X-ray data in both plots are identical, and evidently the warm galaxy LF
  more closely matches the XLF at each redshift.  The XLF was computed
  with redshifts z $<$ 0.5 and z $>$  
  0.5.  
  Also plotted (as open circles) is the local XLF  
  derived in \citet{schmidt1996}, corrected for the factor of $\sim 3$ 
  overdensity of galaxies locally (see text).  
  Curves are plotted for z=0 and the mean redshifts of the two 
  XLF samples ($z=0.27$,$z=0.79$), with the z$>$0 curves being
  calculated assuming (1+z)$^{2.7}$ evolution.  
  The dashed curves show the (minor) 
  effect of including an estimate of X-ray k-corrections when
  converting from FIR to X-ray 0.5-2.0 keV luminosities.  
  The y axis units are log (Mpc$^-3$ dex$^{-1}$).  
  \label{fig:sxlfcomb}} 
\end{figure*}

\begin{figure*} 
% sxlfgalaxy_cdfs_lisa_galaxies_lf60warm_linslope_notitle_04dec03.eps 
\psfig{figure=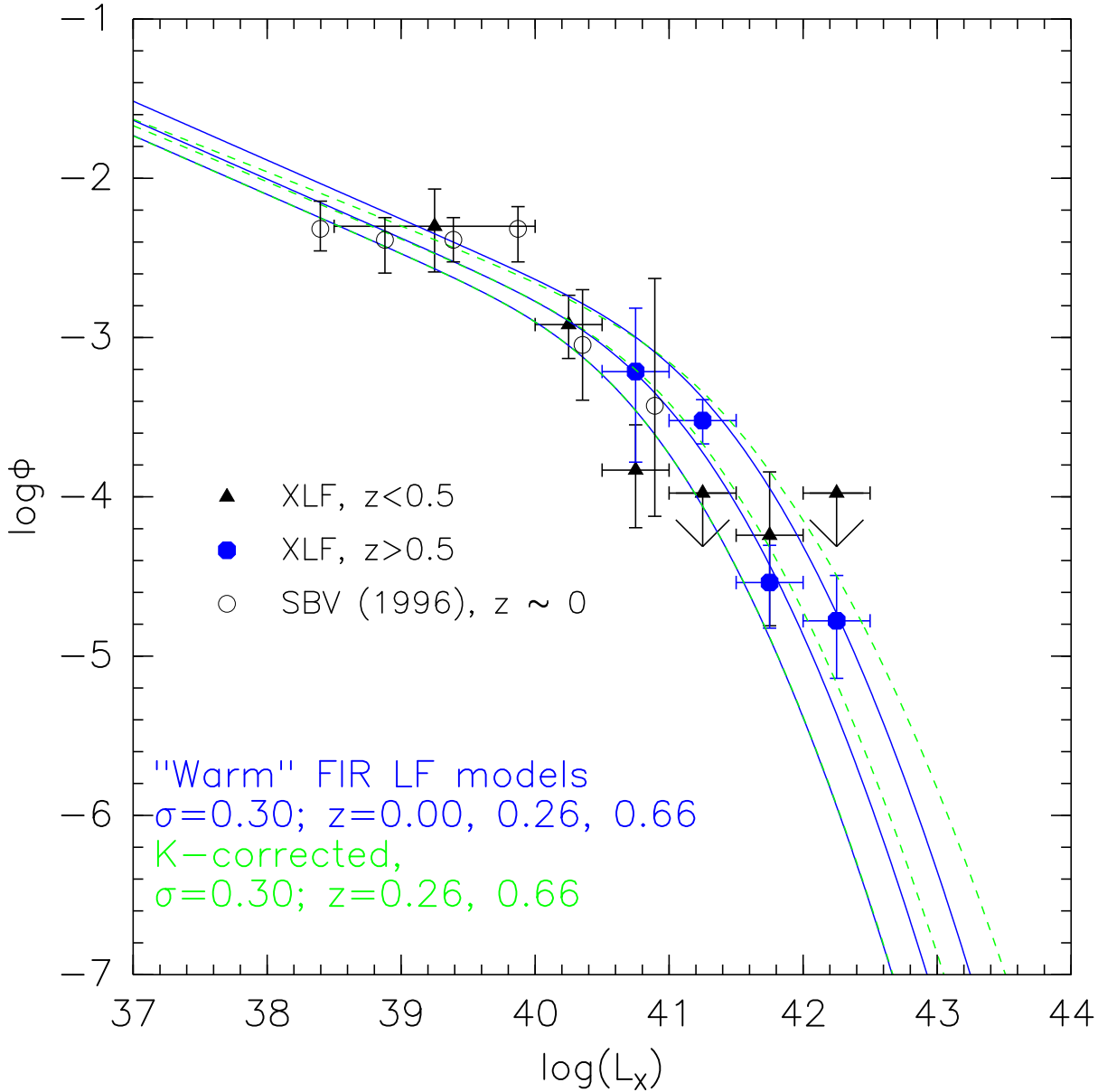,width=16.0cm} 
\figcaption{The CDF-S spectroscopic galaxy sample  (see text for 
  details). Curves and symbols are as in Figure \ref{fig:sxlfcomb}, with the 60
  $\mu m$ ``warm'' galaxy LF model being used here.  The XLF shown 
  here were corrected for completeness, which only impacted the first
  bin of each XLF signficantly (see text).\label{sxlfq3}} 
\end{figure*}

The standard method of constructing binned luminosity functions 
discussed in \citet{page2000} was used in this paper to calculate the 
soft X-ray luminosity function for normal galaxies \citep[see also 
][]{schmidt1968, myaji2000, myaji2001} .  For each redshift and X-ray 
luminosity bin, the source density, $\phi$, can be estimated from 
$\phi \sim N (\int_{L_{min}}^{L_{max}} 
\int_{z_{min}}^{z_{max}(L)}(dV_c/dz)dzdL)^{-1}$, where N is the number 
of sources in the bin, $L_{min}$ and $L_{max}$ are the minimum and 
maximum X-ray luminosities of each bin, and $dV_c/dz$ is the total 
comoving volume per redshift interval, dz, that the CDF surveys can 
reach at each luminosity L.  For a given luminosity L$_X$, 
$z_{min}(L)$ is the minimum redshift of the bin and $z_{max}(L)$ is 
the highest redshift possible for a source of luminosity L for it to 
remain in the redshift bin.  The variance of the source densities can 
also be estimated from $\delta\phi \sim \delta N 
(\int_{L_{min}}^{L_{max}} \int_{z_{min}}^{z_{max}(L)} 
(dV/dz)dzdL)^{-1}$where $\delta$N is the $ 1 \sigma $ Poisson error 
for the number of sources in each bin \citep{kraft1991,gehrels1986}. 
The XLF derived from the Bayesian sample is shown in Figure 
\ref{fig:sxlfcomb} and the spectroscopic CDF-S sample XLF is shown in 
Figure \ref{sxlfq3}.   

In the case of the spectroscopic CDF-S sample 
we also made an approximate correction for spectroscopic completeness. 
%We first computed the fraction of sources in Szokoly et al. (2003) with 
%spectroscopic redshifts as a function of R magnitude.  
We used the mean $\log F_X/F_{opt}$ value for galaxies (-1.7 as 
discussed in the Appedix) to compute the range in R magnitude 
corresponding to each XLF bin, and took the fraction of redshifts in 
that interval (using all sources in Szokoly et al. (2004),  
restricted to a maximum off-axis angle of 8') as an  
estimate of the completeness.  Often in the higher luminosity bins 
there were not a large enough number of sources for the fraction to be 
computed accurately however we found that the completeness when R$<$ 
23 was consistently $\sim 90\%$, and therefore we fixed the 
completeness at 90\% whenever  
the entire higher R range for that bin was  $<$ 23.   
This procedure only significantly 
impacted the first point of each XLF, with completenesses of 43\% and  
73\% for the z$<$0.5 and z$>$0.5 XLFs, respectively. 
 
\subsection{X-ray Luminosity Function Evolution} 
 
\subsubsection{Form of the X-ray Luminosity Function }  
 
To describe the evolution of the X-ray luminosity in more detail it is 
useful to use a functional form for the XLFs we have obtained above. It 
is then easy to test simple models for evolution of the XLFs such as 
pure luminosity evolution (PLE). The choice of the appropriate 
functional fit should, if possible, have a solid basis in physical 
understanding and in empirical observational correlation. After trying 
various forms we chose to use  the lognormal distribution. The 
infrared luminosity function of galaxies (IRLF) is well fit by this 
distribution. In addition, the IRLF is known to be directly 
proportional to the star-formation rate \citep{kennicutt1998}. We 
established that this form of the distribution is a reasonable fit to 
our XLF data. Then, to give the XLF a more physical basis, we {\it 
predict} the XLF from the IRLF based on the \citet{ranalli2003} 
correlations between IR and X-ray fluxes as discussed below. 
  
\subsubsection{IRLF comparison} 
 
We use the 60 $\mu$m luminosity function of \citet{ta03}, 
which assumes the same functional form for the IRLF as 
\citet{saunders1990}, but includes several improvements. Additionally, 
\citet{ta03} adopt the same cosmology used here, allowing for 
direct comparison. We first have to convert to the IRLF to appropriate units 
 to match the XLF. The luminosity function conversions 
were calculated using \citep{georgantopoulos1999, avni1986}: 
 
\begin{equation}
\small
\phi_X (\log{ L_X}) = \int^{+\infty}_{0} \phi_{IR}(\log L_{IR})  
\Phi(\log{ L_X}|\log{ L_{IR}}) d\log{ L_{IR}} 
\normalsize
\end{equation} 
 
where $\Phi(\log{ L_X}|\log{ L_{IR}})$ is the probability distribution 
for observing $L_X$ for a given $L_{IR}$, which we take to be a 
Gaussian, giving  
 
\begin{eqnarray}
\footnotesize
\phi(\log{ L_X}) =  \int^{+\infty}_{-\infty} \phi_{IR}(\log L_{IR})\frac{1}{\sqrt{2\pi}\sigma} \times \nonumber\\
\exp { 
\frac{-(a \log{ L_{IR}} + b - \log L_{X})^2}{\sigma^2}}  d\log{ L_{IR}}
\normalsize
\end{eqnarray} 
 
We calculated the convolution given in eq. 2 numerically  
with the dispersion of $\sigma=0.3$, consistent with the dispersion 
in the soft X-ray / $L_{60 \mu m}$ correlation.   
 
In Figures \ref{fig:sxlfcomb} and \ref{sxlfq3} we show the 60 $\mu$m 
luminosity function of \citet{ta03}, which assumes the same functional 
form for the IRLF as \citet{saunders1990}. In Figure 
\ref{fig:sxlfcomb} we plot the 60 $\mu $m LF from \citet{ta03} 
converted to the 0.5-2.0 keV bandpass using $\log L_{0.5-2.0 \rm \ 
keV} = L_{60 \mu m} - 3.65$ (based on the sample galaxy sample and 
X-ray data as given in Ranalli et al. [2003]). 
 
\begin{figure*} 
\psfig{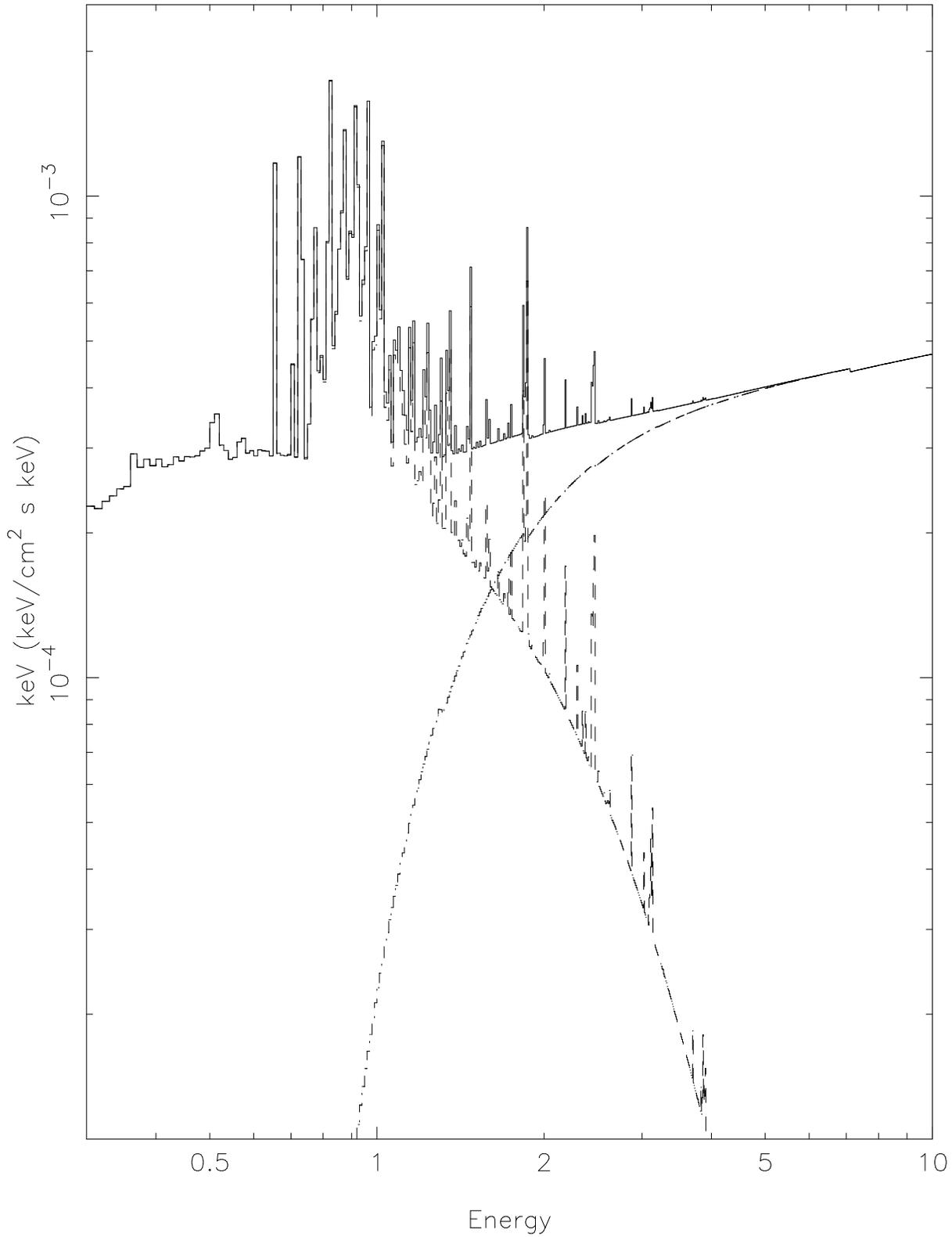} 
\figcaption{Spectral energy distribution starburst model used to 
 ``k-correct'' the models shown in Figures 
  \ref{fig:sxlfcomb} and \ref{sxlfq3}.  
  In the model shown here L(0.5-2.0 keV) = L(2-10 keV), which is the case 
  expected for L(0.5-2.0 keV) = $10^{40} \rm \ ergs \ s^{-1}$ (see 
  text) and corresponds to an effective energy index of 1.0.   The 
  dashed and dot-dashed lines show the contribution of  
  the soft and hard spectral components, respectively. 
\label{fig:sbsed}} 
\end{figure*}

The XLF figures also show the effect of including pure luminosity 
evolution of the form $(1+z)^{2.7}$ (see \S 4) for the mean redshifts 
in each redshift bin (similar to the IR evolution $3 \pm 1$ observed 
in \citet{saunders1990}), and k-correcting the X-ray luminosities 
derived from the IR, i.e., all XLF points are based on luminosities 
computed without k-corrections, equivalent to assuming a flat spectrum 
with an energy index of $\sim 1$, while the IRLF curves where adjusted 
as discussed below.  % (see \S~\ref{sect:kcorr}).   
We assumed a 
canonical SED (shown in Figure \ref{fig:sbsed}; see Ptak et al. 1999) 
with a soft thermal plasma component (kT = 0.7 keV and 0.1 solar 
abundances), representing hot gas possibly associated with superwinds, 
and an absorbed power-law component representing the X-ray binary 
emission ($N_H = 10^{22} \rm \ cm^{-2}$ and an energy index of 0.8). 
The contribution of the hot gas is softer than the binaries but the 
total effect is to give a flat SED.  % whose amplitude correlates well 
with SFR.  The 0.5-2.0 keV/2.0-10.0 keV ratio was varied with 0.5-2.0 
keV luminosity as observed in RCS (going from $\sim 1.7$ at $L_X = 
10^{39} \rm \ ergs \ s^{-1}$ to $\sim 0.3$ at $L_X = 10^{42} \rm \ 
ergs \ s^{-1}$, derived from the $L_{0.5-2.0\rm \ keV}$ and 
$L_{2-10\rm \ keV}$ / FIR correlations). Therefore the typical case is 
in fact consistent with an effective 0.5-10 keV energy index of $\sim 
1.0$, or a $\nu F_{\nu}$ slope of 0. 
  
As a consistency check, we computed the hardness ratios corresponding 
to these model spectra. The model HRs ranged from $-0.8 <HR< -0.3$ 
which can be contrasted with the observed HR $\sim < -0.5$ although 
for the vast majority of our sources there was no significant 
detection in the hard band. In addition, when the observed hard/soft 
X-ray luminosity ratios are correlated with soft X-ray luminosities in 
RCS, the scatter is very large and there is no obvious trend as 
implied by the correlations with FIR.  However, since our assumed 
models are relatively flat, the k-corrections are minor as can be seen 
in Figure \ref{fig:sxlfcomb}. 
 
\citet{ta03} note that the 60 $\mu$m LF binned from the full IRAS PSCz 
sample differs from the LFs derived from ``warm'' and ``cool'' galaxy 
samples, where ``warm'' is defined as $S_{100 \mu m}/S_{60 \mu m} < 
2.1$.  This flux ratio corresponds to a temperature of $\sim 25$K.  In 
Figure \ref{fig:sxlfcomb} we show both the full and warm galaxy sample 
LFs and evidently the XLF is better matched the warm XLF.  The XLF is 
mostly sampling $L > L_{*}$ galaxies, and therefore this is due the 
bright-end LF slope being flatter ($\sigma = 0.625$ as compared to 
$\sigma=0.724$ in the log-normal LF parameterization used here).  This 
may be due to the fact that the galaxies in our sample have relatively 
high implied star formation rates (see below) which results in high 
dust temperatures, however we note that AGN activity will also result 
in higher dust temperatures and X-ray luminosities \citep{miley1985}.

\subsubsection{Radio LF Comparison} 
 
The local radio luminosity function is in excellent 
agreement with the FIR luminosity function \citep{condon2002}, and 
therefore the z=0 FIR model LF is equivalent to the local radio 1.4 
Ghz LF.  This is not surprising considering the tight FIR/1.4 Ghz 
correlation.  
 
\subsubsection{H$\alpha$ Luminosity Function Comparison} 
 
The H$\alpha$ luminosity is a traditional tracer of the massive 
star-formation rate in galaxies \citep{kennicutt1984}. The Balmer line 
strengths are proportional to the number of ionizing photons from 
young massive stars embedded in HII regions \citep{zanstra1927}. Since 
the Balmer lines lie in the red part of the visible spectrum they are 
not as hypersensitive to dust as is the UV continuum. Unlike the FIR, 
H$\alpha$ can be observed to high-redshift. The H$\alpha$ LF is found 
to evolve by up to an order of magnitude by $z=1$ 
\citep{glazebrook1999, yan1999, hopkins2000}. Here we compare the XLF 
to the H$\alpha$ LF since X-rays also are primarily sensitive to the 
massive star formation rate. 
 
We converted the older H$\alpha$ LFs to the now standard cosmology 
adopted for this paper in the following way.  H$\alpha$ luminosity 
functions were calculated assuming $q_{\rm 
 0}=0.5$ and $\Omega_{\Lambda}=0$, with various values for $H_{0}$. 
 Converting for differences in $H_{0}$ is straightforward. However, the 
 cosmological constant introduces a redshift-dependence to the 
 luminosity function translation.   One should return to the 
 original data and re-calculate the luminosity function but  this 
 requires the measured flux values, the redshifts, and detailed 
 knowledge of the sensitivity which are generally not 
 published with the luminosity functions.  As a zeroth-order 
 correction we evaluate the differences in comoving volume and 
 luminosity distance at the median redshift for each luminosity 
 function, and make the appropriate bulk shifts.  We may then compare 
 the H$\alpha$ luminosity functions with our XLF.  For example, the 
 \cite{hopkins2000} H$\alpha$ LF has a median redshift of $z=1.3$. 
 The luminosity distance ($D_L$) for an $\Omega_{\rm M} = 0.3$, 
 $\Omega_{\Lambda}=0.7$, $H_{0}=70$km~s$^{-1}$~Mpc$^{-1}$ Universe is 46\% 
 larger for $z=1.3$ than for the $q_{\rm 0}=0.5$, 
 $H_{0}=75$~km~s$^{-1}$~Mpc$^{-1}$ cosmology of \cite{hopkins2000}. 
 We thus shift their H$\alpha$ LF by a factor of 2.13 to higher 
 luminosity.  The differential comoving volume ($dV_c/dz$) at $z=1.3$ 
 is 3.82 times larger for our cosmology, thus the \cite{hopkins2000} 
 H$\alpha$ LF normalization must be shifted down by this factor. 
 
However, these shifts will not allow for cosmology-dependent changes 
in the {\it shape} of the luminosity function (due to the variance in 
the median redshift of the various bins).  Since we have all the 
required data for the XLF, we thus also calculate our XLF in a 
[$q_{\rm 
0}=0.5$,$\Omega_{\Lambda}=0$,$H_{0}=70$km~s$^{-1}$~Mpc$^{-1}$] 
cosmology to ensure that inferred differences in luminosity functions 
are not just artifacts of our zeroth-order cosmology correction. 
 
\begin{figure*} 
%\psfig{figure=halpha_xlfs_27aug03.eps,width=16.0cm} 
%\psfig{figure=f7.eps,width=16.0cm} 
% halpha_xlfs_10Dec03_H0-70_lamb-0.7.eps 
% halpha_xlfs_08Dec03_H0-70_lamb-0.0.eps 
\plottwo{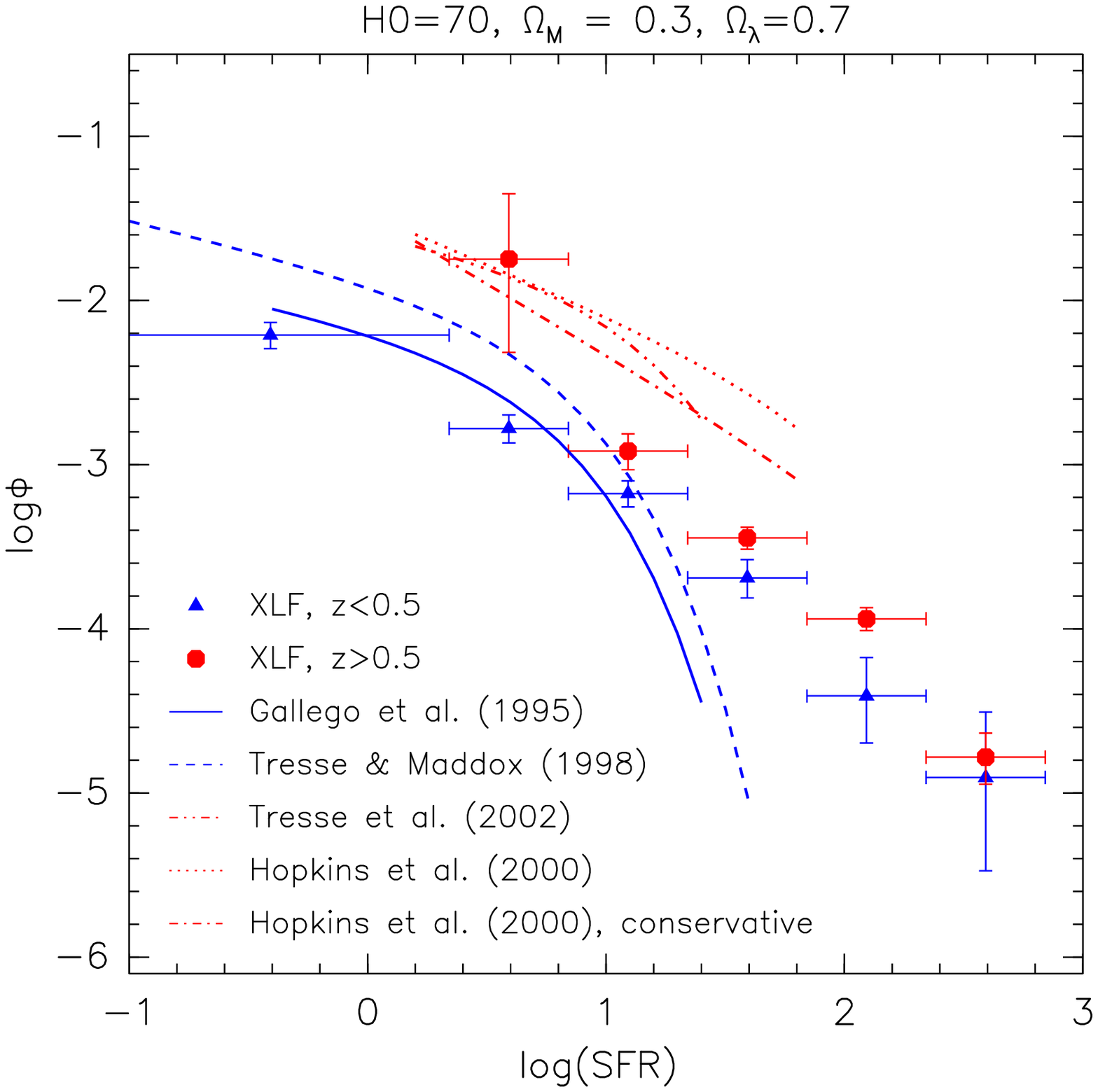}{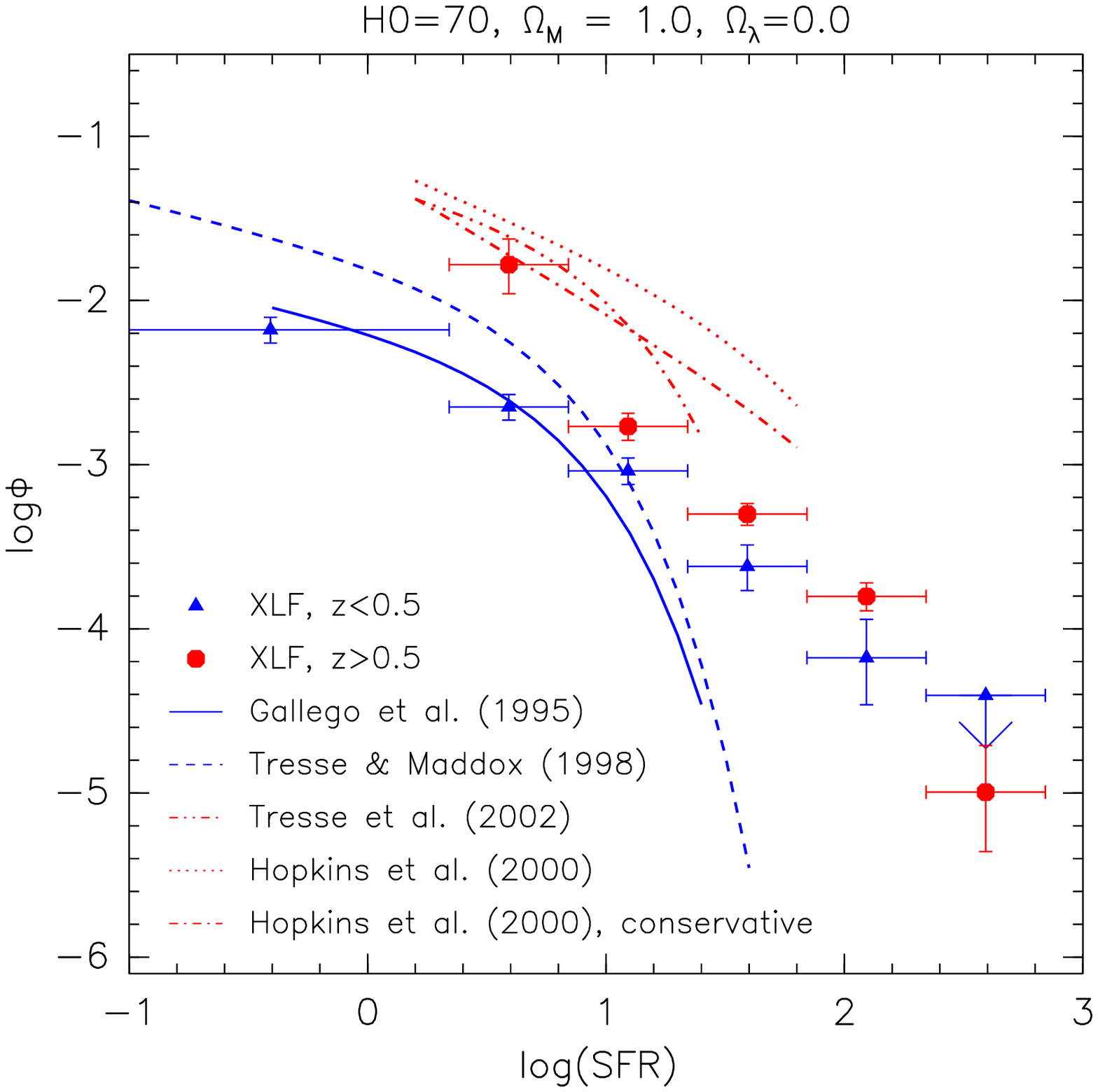} 
\figcaption{ 
Comparison of H$\alpha$ 
and X-ray 
luminosity functions   
(based on the combined CDF-N and CDF-S Bayes-selected sample as shown 
in Figure \ref{fig:sxlfcomb})  
after both have been rescaled to star-formation rates. The H$\alpha$
points were taken from \citet[z $\sim$ 0.7]{tresse2002}, and Schechter
function fits to H$\alpha$ LFs were taken from \citet[z $\sim$
  0.]{gallego1995},\citet[z $\sim$ 0.3]{tresse1998}, and \citet[z $\sim$
  1.3]{hopkins2000}.  Note that \citet{hopkins2000} included the
H$\alpha$ data from \citet{yan1999} in their fits. The H$\alpha$
data are listed in the legend in order of increasing redshift; the z $<$ 0.5 
data and curves are plotted in blue and the z $>$ 0.5 data and curves are 
plotted in red.  The
curves are 
%shown in bold 
plotted only
at the range in luminosity (and hence SFR) over which the fits were performed.
The left panel shows the X-ray luminosity 
function calculated for the cosmology  
$H_0$ = 70 km s$^{-1}$ Mpc$^{-1}$, $\Omega_m = 0.3$ and 
 $\Omega_{\Lambda}=0.7$, and the published $H\alpha$ values  
were converted from the  
cosmologies used 
($H_0$ = 50 km s$^{-1}$ Mpc$^{-1}$, $q_0$ = 0.5 in most cases) by scaling the 
luminosities by $D_L^2$ and the normalization by $dV_c/dz^{-1}$ 
evaluated at the mean redshift of each survey (see text for details). 
The right panel shows the XLF rebinned for a cosmology of $H_0$ = 70 
km s$^{-1}$ Mpc$^{-1}$ and $q_0$ = 0.5, in which case the published 
H$\alpha$ LFs are simply rescaled by $H_0^{-2}$ in luminosity and 
$H_0^{3}$ in normalization.  Both cases are shown here since the
cosmology correction applied in the first case is only an
approximation (out of necessity). \citet{hopkins2000} gave two H$\alpha$ 
  LFs based on a ``conservative'' sample where the H$\alpha$ emission 
  was securely measured and a sample in which less secure H$\alpha$ 
  line identifications were also considered (i.e., the ``true''
  H$\alpha$ LF at this redshift is between these two extremes). 
\label{fig:halpha} 
} 
\end{figure*}

In Figure \ref{fig:halpha} we show H$\alpha$ luminosity functions covering the
full redshift range of interest ($z\approx0$--1.3), for both the
$\Omega_\Lambda =0.7$ and $\Omega_\Lambda=0$ cosmologies.
At lower redshift, the comparison H$\alpha$ LFs include the work of 
\cite{gallego1995} using the Universidad Complutense de Madrid (UCM) 
survey for emission-line objects ($z<0.045$) and the work of 
\cite{tresse1998} using the Canada-France Redshift Survey (CFRS; 
median $z\approx0.2$). At higher redshift, H$\alpha$ shifts into the 
observed near-infrared; the comparison sample at $z\approx0.7$ is the 
VLT ISAAC-observed sample of \cite{tresse2002}.  The extinction 
corrections in \cite{tresse2002} are not as reliable as the 
lower-redshift data due to the difficult of measuring H$\beta$ in the 
NIR spectra.  Finally, at even higher redshift, there are two {\it 
Hubble Space 
  Telescope}    
Near Infrared Camera (NICMOS) studies: \cite{yan1999} and 
\cite{hopkins2000}.  These two studies  
have median $z\approx1.3$ and are not extinction-corrected.

In Figure \ref{fig:halpha}, 
%the H$\alpha$ LF given in 
%\citet{tresse2002} is shown along with the 
the Schechter function fits to the H$\alpha$ LFs from
\citet{gallego1995}, \citet{tresse1998},
\citet{hopkins2000}, and \citet{tresse2002} are shown.  Note that the
results of \citet{yan1999} are 
consistent with those given in \citet{hopkins2000} and were included
in the Schechter function fits in \citet{hopkins2000}.  The H$\alpha$ 
luminosities  are converted to a SFR 
function by multiplying the $H\alpha$ luminosities by $7.9 \times 
10^{-42} \rm M_{\odot} \ yr^{-1} \ ergs^{-1} \ s$ 
\citep{kennicutt1998}.  The XLF based on the combined CDF-S + CDF-N 
Bayes-selected sample shown in Figure \ref{fig:sxlfcomb} was 
converted by multiplying 
the X-ray luminosities by $2.2 \times 10^{-40} \rm M_{\odot} \ yr^{-1} 
\ ergs^{-1} \ s$ (RCS).   Note that the
extinction-corrected z $\sim$ 0.7 H$\alpha$ LF overlaps with the
uncorrected z $\sim$ 1.3 LF, implying that the extinction correction
(a factor of $\sim$ 2) is of the same order as the evolution in the
H$\alpha$ LF between these redshifts.
%Note that the Tresse et al. and Gallego et al. data have 
%been corrected for extinction which is important for properly 
%estimating the SFR using H$\alpha$ luminosities \citep{kewley2002}. 
 
The most striking feature of the XLF / H$\alpha$ LF comparison is that 
the two sets of luminosity functions have different shapes, and more 
specifically different bright-end slopes.  As with the FIR comparison 
the XLF data prefer a flatter bright-end slope than is the case in the 
H$\alpha$ Schecter function fits.  The bright-end of the Schechter 
function is an exponential cut-off, and therefore the only ``tuning'' 
that can be done to fit the bright end of the LF is adjusting $L_{*}$. 
Accordingly the closest agreement  
in shape between the X-ray and H$\alpha$LFs is with the Hopkins et 
al. ``conservative'' Schechter fit, which had the largest value of 
L$_{*}$.  However, note that the shape of the H$\alpha$ LF fits should
only be considered to be representative since the Schechter function
fit parameters are not usually well constrained (and in addition are
often correlated; see Hopkins et al. 2000). 
The relevance of the form
of the luminosity functions is  
discussed below. 
%%%  
%%% In general the reason that the IRLF is fit better with a 
%%% log-normal function rather then Schecter function is that galaxies 
%%% with higher SFR also on average have higher dust masses.  The larger 
%%% amount dust results in higher FIR luminosities while suppressing 
%%% H$\alpha$ emission, resulting in a broader (and hence flatter) FIRLF. 
%%% However, AGN contamination may also be flattening the FIRLF relative 
%%% to the H$\alpha$ LF.  It is likely both the lack of significant extinction 
%%% at keV energies and AGN contamination that is causing the X-ray LF to 
%%% show better agreement with the FIRLF than with the H$\alpha$LF.   
%%%  
 
\subsection{Towards a Physical Understanding of the Luminosity 
Functions associated with Massive Star Formation Indicators.} 
 
The galaxy luminosity functions for cosmic star formation indicators 
have either: (I) a Schechter distribution  or (II) a lognormal 
distribution  and we discuss each of them in turn below.

\subsubsection{Schechter Distribution}. 
 
The Schechter distribution was derived to fit the luminosity function 
for normal galaxies. It is associated with the hierarchical build up 
of luminous baryonic mass, in dark matter potential wells, to form the luminous galaxy distribution function. 
It is obvious that if: (1) the initial stellar mass function(IMF) is approximately 
universal; (2) the (massive) star formation rate per unit luminosity 
is  roughly constant and (3) the H$\alpha$ is not significantly obscured by dust then the 
H$\alpha$LF should roughly reflect the parent Schechter LF of the galaxies.

\subsubsection{Lognormal Distribution} 
 
Lognormal distributions are associated with physical systems that depend on 
many random multiplicative processes (and many associated random multiplicative 
probability distributions). A good example is the complex physics leading to 
relatively simple and robust probability density functions in a 
multi-phase,turbulent self-gravitating system \citep{wada2001}. It is 
not surprising a lognormal distribution results for complex star 
forming systems with,for example, either:(1) reprocessing of the radiation by dust into the 
IR,to make the IRLF or;(2) evolutionary processes of massive stars 
in binaries accreting matter to produce the X-rays of the XLF. One way to think of this is to  
imagine, for example in the x-ray binary emission case, all the time ordered processes that a piece of matter undergoes on its way to emit an X-ray photon from accretion processes onto compact objects in binary systems. Subsequently, the photon may also be reprocessed on its way to the observer. To make this  
argument more explicit let us assume that the Schechter function discussed above  
for the H$\alpha$LF is dependent on the variable X(SH$\alpha$) given by: 
 
\begin{equation} 
X(SH\alpha) = x(star) x(IMF) x(U) x(f_B)= \prod_{i=1}^{4} x(i) 
\end{equation} 
 
where the four x(i) functions on the right are related to star formation, the IMF, the cosmological parameters including $\Omega_m$, and the baryonic fraction respectively. This schematic model is meant to indicate that the H$\alpha$LF depends on a few variables. 
 
The physics behind the production of X-ray luminosity are more complicated. For example for the X-ray luminosity variable X(xray)
we could write:
%\\
%\\
\begin{flushright}
 X(xray) =  X(SH$\alpha$) x(binary) x(compact) x(high mass) x 
(accretion) x(metallicity) x(HI)... = $\prod_{i=1}^{N} x(i) $
\end{flushright}
%\normalsize
%\end{eqnarray} 
%\\
%\\
where here the, N, x(i) functions on the right side depend not only on the four x(i) variables that we have included in our schematic representation of the Schecter  
$H\alpha$ LF but on a total of, N, x(i) functions describing the formation of binaries, compact objects in the binaries, high mass companions, physics of accretion, metallicity and HI column ... respectively. 
Here, the point is that the production of X-rays depends on many more variables, $N>>1$,  in an approximately multiplicative fashion. Consequently, the natural  
distribution for large N multiplicative variables is the lognormal.  
 
Similar ideas apply to the IRLF where, again, we can write: 
%\\
%\\
% 
%\begin{equation} 
\begin{center}
X(IR) = X(SH$\alpha$) x(dust) x(transfer) x(metals) x(SN:dest/form) x(molcloud) x(m,i) x(dyn)...\\
= $\prod_{i=1}^{N} x(i)$
\end{center}
%\end{equation} 
% 
%\\
%\\
where again we have, in addition to the basic Schecter variables, additional x function variables describing the  
physics of dust, radiative transfer, supernovae destruction and formation of grains, the molecular environment, the triggering of star formation by mergers and interactions and other dynamical instabilities (such as bars, and oval distortions).. respectively.   
 
In summary, our argument is that the lognormal distributions, such as the IRLF, XLF and RadioLF are functions of many complicated physical processes, as discussed above, and the Schechter type functions, such as the H$\alpha$LF and the K-band LF depend on relatively simple and relatively few processes.

\subsubsection{Future Work} 
 
Detailed models could be developed using the ideas discussed above but 
with more attention to the detailed physical processes. Possibly, the 
lognormal distributions may be associated more with starbursting 
normal galaxies rather than normal quiescent galaxies. It is likely 
that starbursting events are triggered in some way, by external galaxy 
merging and interactions or by internal dynamical instabilities such 
as bar formation. The random nature of such triggering processes may 
naturally lead to the type of 
 multiplicative probability chain that would produce  
lognormal distributions. The more 
normal star formation mode may be reflected by the H$\alpha$. It 
would then be very interesting that normal modes and the triggered 
modes contribute very roughly the same amount of the cosmic SFR. For a 
detailed discussion of the merits of H$\alpha$ {\it versus} IR star 
formation rate indicators see \citet{kewley2002}. 
 
\section{Cosmic Star Formation History} 
 
\begin{figure*} 
\plotone{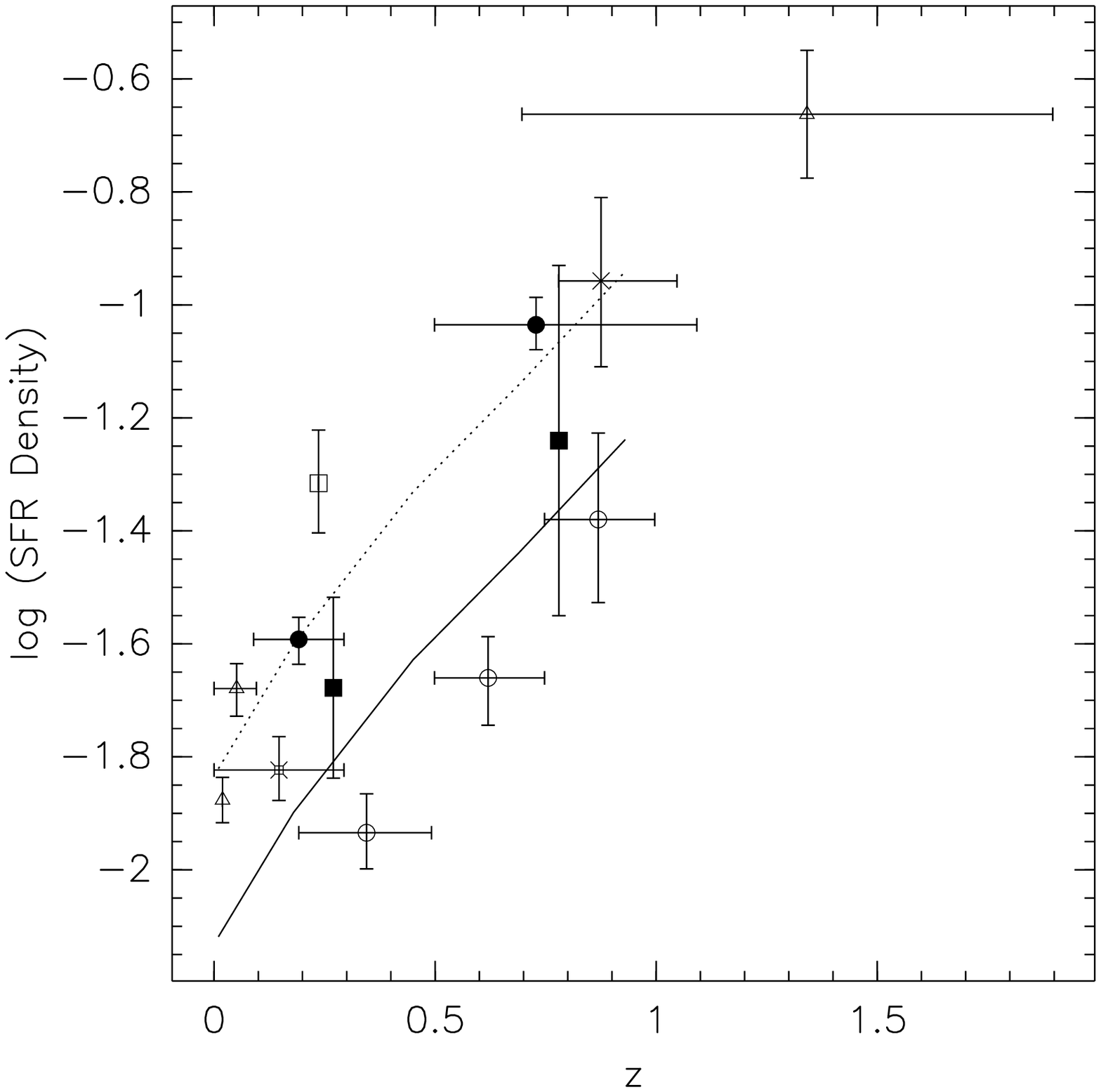} 
\figcaption{Compilation of SFR densities from \citet{tresse2002}, including 
  the X-ray points for z=0.27 and z=0.79 from this work.  The X-ray 
  points are shown with black filled squares. The triangles represent
  the H$\alpha$ SFR values from \citet{gallego1995} at z $\sim 0$,
  \citet{gronwall1999} at z$\sim 0.05$, and \citet{hopkins2000} at z$\sim
  1.3$.  The empty square represents the UV-selected H$\alpha$
  \citet{pascual2001} value at z$\sim 
  0.24$.  The filled circles show the SFR densities from
  \citet{tresse2002}. The star gives the UV-selected z=0.15 SFR density from
  \citet{sullivan2000}.  The open circles give 2800\AA~ CFRS points
  from \citet{lilly1996}, without dust correction. We also plot the
  SFR history based on the 60 $\mu$m LF luminosity density at z $\sim
  0$, including (1+z)$^3$ evolution.  The 60 $\mu$m luminosity density
  was computed from \citet[solid line]{saunders1990} and \citet[dotted
    line]{ta03}.
\label{fig:sfrplot}} 
\end{figure*}

In Figure \ref{fig:sfrplot} we show the cosmic star formation history 
taken from \citet{tresse2002}, with X-ray points added at z=0.27 and z=0.79 
based on the combined CDF-N+S Bayes sample.  We chose the Bayes sample
since the XLF is in better agreement with the FIR prediction (see 
below).  
We computed the X-ray luminosity density by integrating $\Phi 
(logL) L dlogL$ and rescaling by  
$2.2 \times 10^{-40} \rm M_{\odot} \ yr^{-1} 
\ erg\ s^{-1}$ (RCS).  
This resulted in SFR density estimates of $2.0\times 10^{-2}$ and  
$6.0\times 10^{-2}\ \rm M_{\odot} \ yr^{-1} Mpc^{-3}$ at two redshifts.
%resulting in SFR of 0.019 and 
%0.024 $M_{\odot} \ yr^{-1} Mpc{^-3}$ at z=0.27 and z=0.79.The X-ray SFR points are shown as filled black squares.     
 
Note that 
the integration of an LF tends to be dominated by the contribution of 
the LF near L$_{*}$.  Here we primarily only sample galaxies 
brighter than L$_{*}$, and therefore 
this estimate is a lower limit to the true SFR.  However we  
also integrated the rescaled FIR LF (the curves in Figure 
\ref{fig:sxlfcomb}b including k-corrections) at  
those redshifts, resulting in SFR = 0.022 and 0.055  
$\rm M_{\odot} \ yr^{-1} Mpc^{-3}$.  
The two approaches resulted in consistent SFR estimates however the 
errors are large on the XLF data points, particularly on the lowest
luminosity points which dominate the  
luminosity density sum.  Therefore, we estimated the error on the SFR
by summing the upper and lower error bounds on the XLF data
points.  This resulted in errors in log SFR of $\sim 0.16$ and $\sim
0.31$ at z=0.27 and 0.79, respectively.  In this figure the SFR density  
estimates from \citet{tresse2002} were corrected to $h = 0.7,
\Omega_{\lambda} = 0.7,  
\Omega_{M}=0.3$.  

The X-ray SFR points are consistent with 
 an evolution of the SFR for $0<z<1$ of SFR $\propto 
(1+z)^{2.7}$ (i.e., an increase by a factor of $\sim 2.5$ in SFR
 density between z $\sim 0.3$ and z $\sim 0.8$). 
The Tresse et al. SFR
 history plot includes UV and extinction-corrected H$\alpha$ values.
 The z $\sim 0.3$ XLF SFR estimate is more consistent with the
 H$\alpha$ SFR estimate and the z $\sim 0.8$ XLF SFR estimate
 is intermediate between the H$\alpha$ and UV SFR values.  However
 since the error on the z $\sim 0.8$ XLF SFR is large it is consistent
 with either the UV or H$\alpha$ SFR.  We also plotted the z $\sim$ 0
 SFR predicted from the \citet{saunders1990} 60 $\mu$m LF (solid line)
 with an  evolution of $(1+z)^3$, as
 well as the 60 $\mu$m  SFR predicted by integrating the 60 $\mu$m LF
 in \citet[dotted
 line]{ta03}.  In both cases the 60 $\mu$m/SFR conversion of 
$2.6 \times 10^{-10} \ M_{\odot} \ \rm yr^{-1} [L_{60}/L_{\odot}]^{-1}$ 
(where $L_{60}/L_{\odot}$ is the 60 $\mu$m luminosity in solar
 units) from
 \citet{rowanrobinson1997} was used.
The \citet{ta03} 60 $\mu$m LF has a somewhat steeper
 faint-end slope, which resulted in a factor of $\sim 2$ larger
 luminosity density.  
%There is, however, obviously a very large amount 
%of scatter between the various methods of determining the SFR history. 
%There is reasonable agreement in the amplitude, the shape and the 
%evolution with redshift of the XLF with the FIR LF and HALF under the 
%assumption that all trace the SFR giving additional confidence that we 
%have selected star-forming galaxies. 
 
\section{Discussion} 
 
Our studies have focused on establishing the X-ray derived cosmic star 
formation rate and the normal star-forming galaxy X-ray Luminosity 
functions up to redshifts of order unity. Mild evolution is seen. The 
results presented here open up a new waveband for such star formation 
studies with the significant advantage that obscuration effects are 
not present. The Bayesian techniques developed in this paper for the 
galaxy selection have a wide range of applicability for further such 
studies. We now discuss important aspects of our results in more 
detail below. 

\subsection{Normal Star-Forming Galaxy X-ray Luminosity Functions} 
 
We have derived the first X-ray {\it galaxy} luminosity functions at 
the redshifts of $\sim 0.3$ and $\sim 0.7$.   
%In all XLF samples 
%discussed above, the high-redshift bin at $\log L_{X} = 40.5$ lies 
%significantly below the rescaled FIR LF.  Since this bin represents 
%the faintest sources in the sample, the flattening of the XLF 
%exhibited in Figures \ref{fig:sxlfcomb} and \ref{sxlfq3} is most 
%likely due to incompleteness, specifically spectroscopic 
%incompleteness.   
The high-luminosity bins for both the z$<$0.5 and z$>$0.5 
Bayesian sample XLFs tend to be flatter than the predictions based on 
rescaling the FIR LF,  
although there is better agreement when the XLF is compared to the 
``warm'' galaxy FIR LF rather than the full PSCz sample FIR LF. 
This implies that there is some AGN contamination in the sample. 
%particularly for the high redshift spectroscopic %sample at high 
%luminosities and  
The spectroscopic CDF-S sample XLF also appears to be somewhat flat, 
which may indicate that absorbed and/or low-luminosity AGN  
are being missed in the optical spectra, although the small number of 
galaxies in that sample results in large statistical errors. Near IR 
spectra may be more  
revealing \citep[e.g.,][]{veilleux99}.  %The problem of AGN 
%contamination is evidently more acute in the CDF-N sample (see Figure 
%\ref{sxlfq3}).   
%In contrast, the Bayes-selected galaxy XLF is in 
%fairly good agreement with the rescaled ``warm'' galaxy FIR LF. % for z$>$0.5 and $\log L_{X} \geq 41$.   
This overall agreement with the FIR prediction suggests that X-rays 
may be a useful tool for  
discriminating AGN2 and galaxies in surveys, due to the ability of 
X-rays to penetrate columns of $10^{22-23} \ \rm cm^{-2}$ as typically 
observed in Compton-thin AGN2.  Finally, $\log L_{X} < 40.5$ bin from 
the Bayesian z $<$ 0.5 XLF  
% and the corresponding spectroscopic 
%$<$ 0.5 XLF bin prior to completeness correction are 
is most consistent 
with the local  
(full sample) FIR LF and the z=0 XLF from \citet{schmidt1996}, implying no 
evolution.  However the mean redshift of galaxies with $\log L_{X} < 
40.5$ is only 0.16, implying evolution only on the order of $\sim 
50\%$ in luminosity. 
 
\subsection{Star Formation Rates from X-ray Studies} 
 
Detailed physical modeling of the X-ray emission from galaxies and its 
application in measuring cosmic star formation history has been 
considered recently by \citet{cavaliere2000} for the hot gas 
component, \citet{ghosh2001} and \citet{ptak2001} for the X-ray binary 
component, and for high mass binaries by \citet{grimm2003}. The 
spectral energy distribution (SED) for starburst galaxies has been 
modeled by \citet{persic2002}. The potentially dominant contribution 
from Ultra-Luminous X-ray point sources (ULXs) has been analyzed in a 
recent paper by \citet{colbert2003}.   
 
The physical reason for the correlation of both hard and soft X-ray 
luminosity with SFR is worth studying because it seems to constrain the  
overall physics of massive star formation and evolution.  
One possibility is that the (accretion-powered) 
core-collapse supernovae provide the hot $1$ keV component and the 
accretion-powered high-mass X-ray binaries dominating the point source 
contribution \citep{grimm2003}. However, the detailed physics of how 
the relative branching ratios to SNe and to HMXBs for the evolution of 
the massive star population can conspire to produce the flat SED is 
not yet clear. We argue below that not only must the single star IMF be fixed  
but the bivariate binary star IMF($M_1, M_2$) must also be approximately constant. 
For a flat SED, what is required is that the massive stars ending in Type II SNe powering the 1 keV hot gas component of the ISM 
and the HMXBs for the evolution of massive stars in binaries must be giving approximately equal energy per octave.   
Detailed studies of nearby galaxies and detailed 
population synthesis models help significantly \citep{zezas2002, 
georgantopoulos2003, swartz2003, colbert2003, sipior2003}. 
 In general, the X-ray luminosity of a galaxy can be related 
to the mass of the galaxy, $M$ and its star formation rate, $SFR$ by : 
 
\begin{equation} 
L_{X} = \eta_M M_G + \eta_* {\dot M_*} 
\end{equation} 
 
where the $\eta$'s are linear efficiency factors. The first term is 
due to the old low mass X-ray binary population(LMXBs) and the second 
term is due to the high mass binaries (HMXBs) and the hot gas powered 
by SNe. In the {\it Chandra} band (0.3-8 keV), the X-ray point-source 
luminosity is proportional to the mass and SFR with coefficients 
$\eta_M = 1.3 \times 10^{29} \rm \ ergs 
\ s^{-1} \ M_{\odot}^{-1}$ and 
$\eta_* = 0.7 \times 10^{39} \rm \ ergs \ s^{-1} \ (M_{\odot} \ 
yr^{-1})^{-1}$ \citep{colbert2003}. In the soft band (0.5-2.0 keV) the 
star formation dominates for galaxies with $L_{X} \ga 10^{39} \ \rm \ 
ergs 
\ s^{-1}$ (as indicated in X-ray observations of the composite 
emission of starburst galaxies; \citet{dahlem1998}; \citet{ptak1999}). 
It was, therefore, not surprising that \citet{ranalli2003} found that 
the soft X-ray luminosity of starburst galaxies is proportional to the 
SFR alone, specifically with $\eta_M \sim 0$ and $\eta_* = 4.5 \times 10^{39}\rm \ 
ergs \ s^{-1} \ (M_{\odot} \ yr^{-1})^{-1}$.

The extinction-corrected H$\alpha$ star formation history and the 
X-ray star formation history formally agree within the errors,
although the X-ray SFR values are lower than the H$\alpha$ SFR
values. This is probably due in part to the different sample selection
and the different corrections applied to calculate the SFR.  The
revision to the 60 $\mu$m LF given in \citet{ta03} resulted in a factor
of $\sim 2$ increase in the 60 luminosity density relative to the
value given in \citet{saunders1990}.  It is likely that future work on
both the H$\alpha$ LF and XLF will similarly result in adjustments to
the SFR estimates.
 
The recent comprehensive study of \citet{cohen2003} using spectroscopy 
of the [OII]3727 $\AA$ line emitted by galaxies found in the 2 Ms 
exposure of the CDF-N shows that the $L_X$-SFR rate is similar locally 
and at redshifts of order unity, which is consistent with what we 
assume here. 
 
The basic ingredients of a workable model are fairly clear: 
 
(1)Constant IMF($M_1, M_2$). 
 
 A constant two-parameter IMF(B($M_1$, $M_2$),S($M_1$))for binary stars(B)  
and single stars(S) as a function of $M_1$ and $M_2$. This is probably an important implication  
of the work of \citet{ranalli2003} and if the general consistency here between the x-ray star formation results 
and the star formation indicators. 
 
(2)Normal Stellar Evolution and Binary Star Evolution. 
 
 Straightforward stellar evolution and binary evolution to produce roughly constant  
branching ratios for high-mass X-ray binaries (HMXB) and core collapse SNeII. This is probably 
 satisfied since we expect no surprises in stellar evolution and binary evolution \citep{ghosh2001}. 
  
 (3)Consistent ISM properties. 
 
It is also necessary to have at least very roughly constant properties for the interstellar medium(ISM). 
Over the whole galaxy ensemble, no large variations in the properties of the ISM 
including: dust-to-gas ratio, grain properties, metallicity, dense molecular environments and galactic winds. However, the dispersion of ISM properties  from galaxy-to-galaxy 
is known to be large. For example, there are large well-known variations in the grain properties in the LMC, the SMC and the Galaxy. 
 
(4)Consistent Dynamical Properties.   
 
The population of normal starbursting galaxies we are studying can have additional dynamical and environmental  
constraints due to the amount of merging, interactions, barred driven inflow...  
 
The first two points (1,2) give a way to understand the rough correlations and the last two points (3,4) give the dispersion and the 
 overall lognormal shape to the radio, IR and X-ray LFs.  
 
The star formation history derived from the X-ray data using the 
empirically derived relation between X-ray luminosity and star 
formation rate from \citet{ranalli2003} is in rough agreement with other 
methods although, as discussed by \citet{hogg2004} there is a large dispersion in the results.

\subsection{Hidden Active Galactic Nuclei} 
 
An important issue is whether the X-ray emission from our galaxies 
could be due to central AGNs hidden in faint Type 2 Seyfert nuclei 
lurking in the center of our galaxies \citet{moran2002}. This is not 
likely because Seyfert 2s have a hard spectrum with hardness ratio $HR 
> 0$, whereas the galaxies in our sample have softer spectra. 
 
A key advantage to using soft X-ray luminosities to estimate the X-ray 
galaxy luminosity function and the SFR is that the soft band tends to be 
dominated by starburst emission {\it regardless of whether an 
obscured AGN is present} \citep[see, e.g.,][]{turner1997, 
levenson2001}.  This assists us in discerning galaxies 
from AGN2. 
 
\subsection{Implications for Future X-ray Missions} 
 
Our cosmological star formation studies, extending up to $z\approx1$, 
require the combination of 1-2 Msec $Chandra$ exposures and high-quality 
optical spectroscopic coverage from large-aperture ground-based telescopes 
such as VLT and Keck.  There is, however, a wealth of information on 
galaxy evolution available through current ground-based wide-field optical 
surveys (e.g., 2dF and SDSS) that is still relatively underutilized. The 
small fields of view of $Chandra$ and $XMM$-$Newton$ make X-ray studies of 
these wide-field galaxy surveys quiet difficult \citep[but see e.g.,][for 
an example of one high-quality {\it statistical} study of 2dF galaxies 
with $XMM$-$Newton$]{georgakakis2003}.  Deeper wide-field X-ray surveys 
would also be complementary to the deep wide-field ultraviolet imaging 
surveys of GALEX. The GALEX surveys, currently underway, are probing star 
formation from $z=0-2$, overlapping with the work we present here. 
Possible future wide-field X-ray telescopes include the Wide Field X-ray 
Telescope (WFXT) \citet{burrows1992} and its successors WAXS/WFXT 
\citep{chincarini1998} and DUET (Jahoda et al. 2003). 
 
X-ray studies of galaxies may be extended to truly extreme redshifts when 
high-sensitivity X-ray telescopes such as XEUS \citep{Parmar2003} commence 
operation.  XEUS may achieve sensitivities as faint as 
$4\times10^{-18}$~erg~cm$^{-1}$ and at this level galaxies will dominate 
the X-ray source number counts.  This sensitivity will allow for the 
detection of star-forming galaxies to great distances ($z>5$) and will 
allow for the probing of X-ray emission from star formation to the very 
epoch of galaxy formation in the early Universe. 
 
\section{Conclusions} 
 
There are five main conclusions of this work. 
 
1. We have established the X-ray luminosity functions of galaxies in 
the {\it Chandra} deep Fields North and South. These XLFs have a 
lognormal distribution. The XLFs are completely consistent with both 
the infra-red luminosity functions (IRLFs) and the GHz radio luminosity 
functions(RLFs), which both have lognormal distributions. 
 
2. The evolution of the XLFs is consistent with a pure luminosity 
evolution(PLE) of $\sim (1+z)^{2.7}$.The more robust integral of the 
XLFs can be transformed to an x-ray derived cosmic star formation 
history. This cosmic star formation rate (SFR) is consistent with the 
(mild) evolution in SFR $\propto (1 + z)^{2.7}$. In general, 
the overall SFR estimates in the range of redshift from unity to the 
present epoch show a wide scatter although there is a general trend of 
mild evolution consistent with the above. To reduce the scatter and 
better understand the evolution of the SFR in this redshift range one 
needs in the future:(1) more sensitive wide area studies such as the 
GALEX mission (2) extensive multi-band studies; the XLFs and X-ray 
derived SFRs now bring in a new waveband here and (3) deeper physical 
understanding of the star formation indicators.

3. The H$\alpha$LFs have the form of the Schechter luminosity 
function which fits the galaxy luminosity functions in J and K. The H 
$\alpha$ distribution is quite different from the the lognormal 
distributions for the XLF, IRLF and RLF discussed above. A detailed 
physical understanding is not yet available.

4. The X-ray derived star formation history gives an interesting constraint on the IMF. 
The X-ray emission in the Chandra band is most likely due to binary stars. The overall consistency and correlations between single star effects and binary star effects 
in different wavebands indicate that the bivariate IMF($M_1, M_2$) must be universally constant, at least at the high mass end, since the  X-ray observations may be measuring directly the {\bf \it binary star formation history} of the Universe.

\acknowledgements 
 
It is a pleasure to thank both the CXO Director, Harvey Tannenbaum, 
and the CXO team and the ESO Director, Catherine Cesarsky, and the ESO 
staff for their joint heroic efforts that made this project possible. 
We thank David Alexander for useful advice concerning the CDF X-ray 
data analysis and for sharing data.  We also thank David Hogg for sharing his 
SFR compilation data with us.  We thank Ivan Baldry for useful discussions 
concerning this work.
 
\appendix 
 
\section{ Bayesian Statistical Analysis} 
 
For many of the sources in our sample both the X-ray and optical 
information is limited.  Classifying a source based simply on 
``sharp'' cuts in parameter space, such as log $L_X < 42 \ \rm ergs \ 
s^{-1}$ and X-ray hardness HR $< -0.2$, does not take into account the 
errors in the measured quantities and the fact that the parent 
distribution (i.e., real galaxies in the Universe) are not so sharply 
delineated.  Here we remedy this situation by using a Bayesian 
statistical approach.  Specifically, we assume parent distribution 
models for measured parameters, $\theta$, and then compute the 
conditional probability, $P_M$, of measuring the observed values of 
the parameters ($D$) by integrating over the parent distributions for 
a given model M. Specifically, $\theta$ = \{HR, $\log L_X$, $\log 
F_X/F_{opt}$\}, where HR is the X-ray hardness ratio, $F_X$ is the 
0.5-2.0 keV luminosity, and $\log F_X/F_{opt}$ is the X-ray/optical 
flux ratio.  $\log F_X/F_{opt}$ is given by $\log_{10}(F_{0.5-2.0 \rm \ 
  keV} - R/2.5 - R_0$, where $R_0$ = 5.5 for CDF-N sources 
\citep{hornschemeier2003} and $R_0$ = 5.7 for CDF-S sources 
\citep{szokoly2004}. 
$P_M$ is given by $P_M = \int p_M(\theta|D) d\theta$, with 
\begin{equation} 
%P(L, H) = \int dL' \int dH' P_M(L', H') {\mathcal L}(L | L') 
%{\mathcal L}(H | H') 
p_M(\theta|D) = p_M(\theta) p(D|\theta) / p(D) 
\end{equation} 
$p(D|\theta)/p(D)$ is the probability of observing a given set of 
parameter values D when the expected (i.e., parent) values are  
$\theta$, which is simply the likelihood function for the 
data. In this analysis we assume that the errors on each parameter are 
reasonably approximately Gaussian, and defer the treatment of the 
more general likelihood functions for future work.   
$p_M$ is the prior, or parent population probability distribution for 
a sources of type M.  We further assume that the parent distributions 
of the parameters are  
also reasonably approximated by a Gaussian, with the mean and standard 
deviations of the Gaussian set to the observed values for the 
parameter (see below).  $p_M$ should be normalized 
by the relative number of sources of a given type that would be 
detected in the 0.5-2.0 keV bandpass.  However, since these numbers 
are not known precisely and the relative number of sources used in the 
statistical analysis below are approximately equal, we set this term 
to unity here. 
Therefore $P_M$ is given by 
\begin{equation} 
P_M = \Pi_i \int d\theta_i \frac{1}{\sqrt{2\pi}\sigma_i} 
  \exp[-\frac{(\theta_i - \bar{\theta_i})^2}{2\sigma_i^2}] 
  \frac{1}{\sqrt{2\pi}\Delta x_i}\exp[ -\frac{(x_i - 
  \theta_i)^2}{2\Delta x_i^2}] 
\end{equation} 
Here $\bar \theta_i$ and $\sigma_{i}$ are the parent distribution mean 
and standard deviation for the ith parameter, while $x_i$ and $\Delta 
x_i$ are the measured value and error for that parameter.  
%Note that Equation A1 is also referred to as   
%Bayesian evidence by some groups \citep{hobson2003}. 
For more discussion of Bayesian methods used in model testing, see 
\citet{hobson2003} and references therein. 
 
\begin{figure} 
\plotone{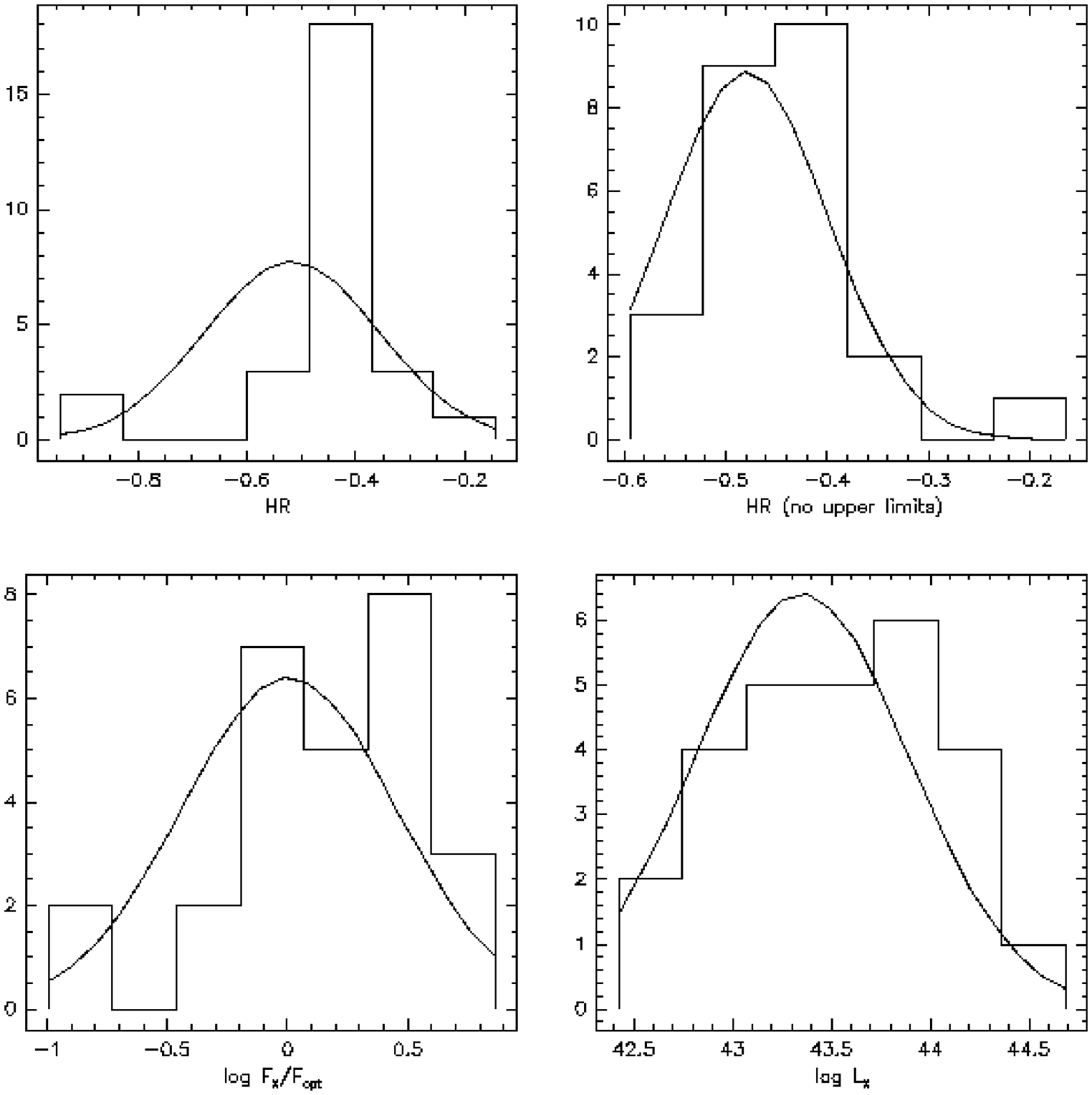} 
\figcaption{The observed parameter distributions for AGN1 sources in 
  the CDF-S.  The y-axes are numbers of sources.  The curves show the 
  Gaussian resulting from taking the mean and standard deviation of 
  the distribution. \label{agn1fig}} 
\end{figure}

\begin{figure} 
\plotone{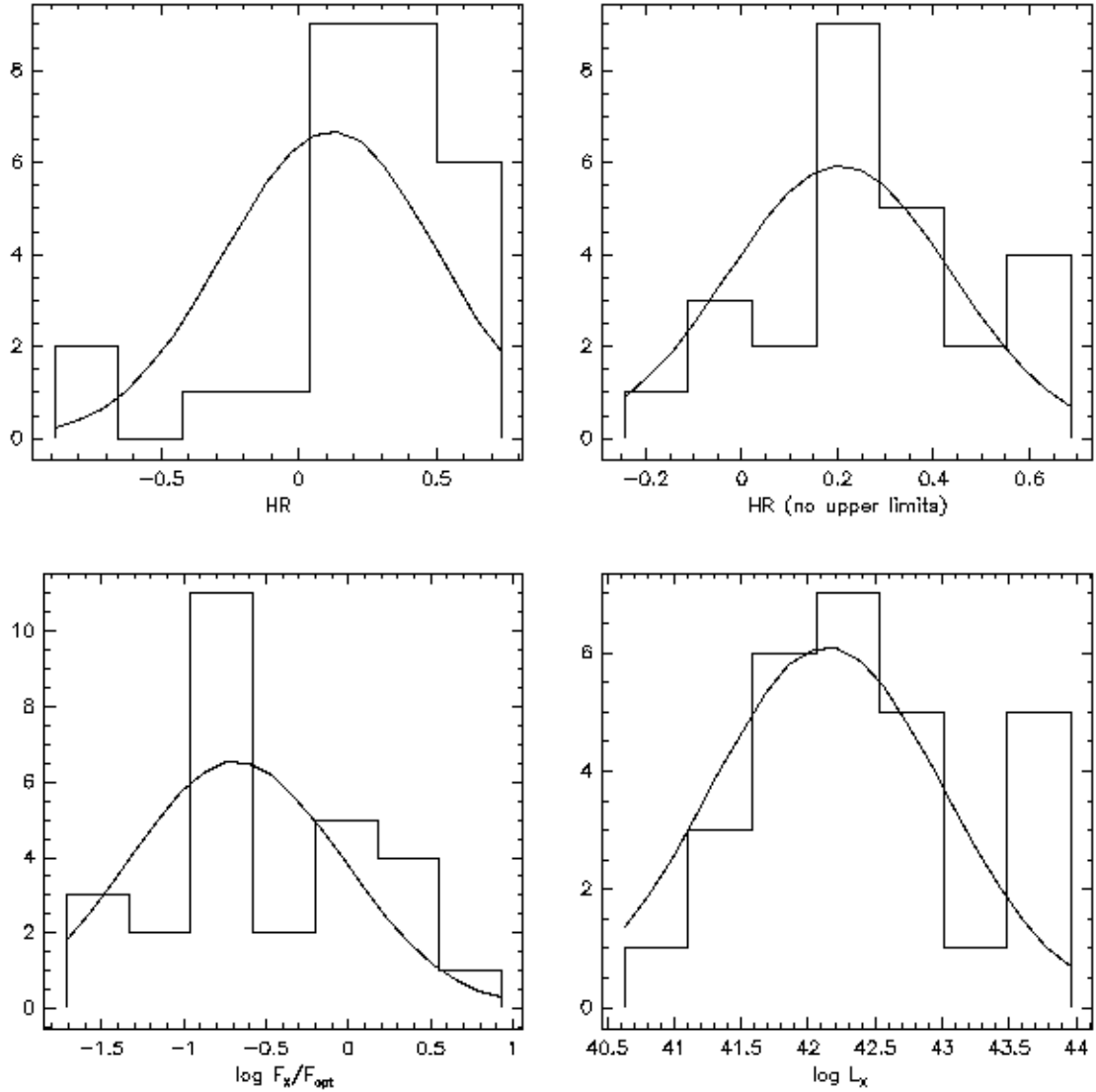} 
\figcaption{As in Figure \ref{agn1fig}, except for AGN2 
  sources. \label{agn2fig}}  
\end{figure}

\begin{figure} 
\plotone{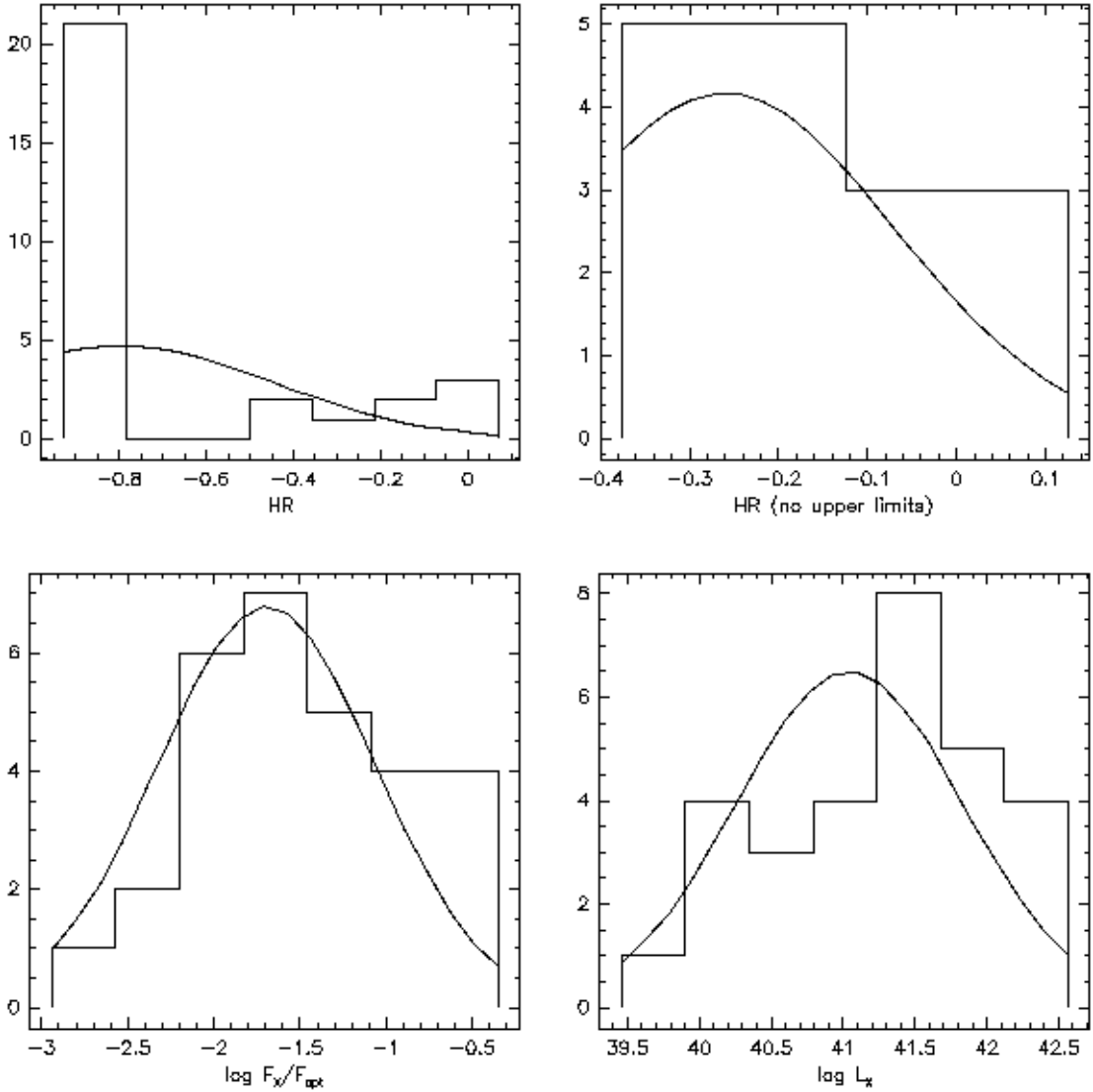} 
\figcaption{As in Figure \ref{agn1fig}, except for galaxy 
  sources. \label{galfig}}  
\end{figure}

Applying these methods to our sample, we have chosen Gaussian parent 
distribution functions whose mean and standard deviations ($\sigma$) 
are derived from the subset of X-ray sources which have high-quality 
optical spectroscopic identifications 
\citep{szokoly2004}\footnote{Ambiguous cases with $L_X > 
 10^{42.5}\ \rm ergs \ s^{-1}$ where classified as AGN1 galaxies}.  We further 
 require that the sources have  
 constrained hardness ratios (relevant mostly for the galaxies, which were 
 X-ray faint) when computing the mean and standard deviation of the 
 hardness ratios. The histograms of the observed parameters are shown 
 in Figures \ref{agn1fig}-\ref{galfig}.  The Gaussian parameters are 
 shown in Table \ref{app:modelparams} and the corresponding Gaussian 
 curve is also plotted in Figures \ref{agn1fig}-\ref{galfig}.  

\begin{deluxetable}{llllllll} 
\tablecaption{Statistical Parameters for the Galaxy, AGN1 and 
  AGN2 CDF-S Samples \label{app:modelparams}} 
\tablehead{ 
\colhead{Source Type} & 
 \colhead{Number} & 
 \colhead{$\bar{\log L_X}$} & 
\colhead{$\sigma_{\log L_X}$} &  
\colhead{$\bar{HR}$} & 
\colhead{$\sigma_{HR}$}  
& \colhead{$\bar{\log F_X/F_{opt}}$} & 
\colhead{$\sigma_{\log F_X/F_{opt}}$} 
} 
\startdata 
%Galaxy & 31  &  40.8 & 0.9 & -0.35 & 0.29 \\ 
% Submitted: 
%Galaxy & 29  &  41.0 & 0.8 & -0.26 & 0.19 \\ 
%AGN1 &  24 &  43.4 & 0.6 & -0.49 & 0.05 \\ 
%AGN2 &  29  & 42.2 & 0.9 & 0.13 & 0.31 \\ 
Galaxy & 8/29 & 41.0 & 0.8 & -0.26 & 0.19 & -1.7 & 0.6 \\ 
AGN1 &  25/27 & 43.4 & 0.5 & -0.48 & 0.08 & 0.0 & 0.4 \\ 
AGN2 &  26/28  & 42.1 & 0.9 & 0.21 & 0.23 & -0.7 & 0.6 \\ 
\enddata 
\tablecomments{The Number column lists the number of sources of the 
  given type with a detection in the hard band / the total number of 
  sources of that type.  Sources with upper-limits on the hardness 
  ratio (i.e., without hard band detections) were excluded when 
  computing the mean and standard deviation of HR.}    
\end{deluxetable} 
 
Using these assumed models, the observed errors on HR, and an assumed 
error of 0.25 on $\log L_X$ and $\log F_X/F_{opt}$, we computed the 
probability of each source being a galaxy, an AGN1 and an AGN2.  Note 
that the errors on HR were also taken to be Gaussian, using the larger 
of the upper and lower error estimates \citep[derived using][; see 
discussion in Alexander et al. 2003]{lyons91,kraft1991}.  A source is 
classified as a given type when the computed probability for that type 
is the greatest.  As a check, we summarize in Table \ref{bayescheck} 
the results for the sources with high-quality classifications (i.e., 
those used to produce the parent distribution models as discussed 
above).  Of course in general a control sample is necessary to 
properly assess the performance of a classification procedure, however 
our sample is too small to allow for this approach.  However in future 
work this test will be performed with other data sets (most notably 
using GOODS data).  As would be expected, %from Figure 
nearly all of the false galaxy identifications were 
AGN2 and vice-versa. 
 
\begin{deluxetable}{llllll} 
\tablecaption{Bayesian Classification Results \label{bayescheck}} 
\tablehead{ 
\colhead{Source Type} & \colhead{Number} & 
\colhead{Galaxy} & \colhead{AGN1} & \colhead{AGN2}} 
\startdata 
Galaxy & 29 & 25 & 2 & 2\\ 
AGN1 & 27 & 1 & 25 & 1 & \\ 
AGN2 & 28 & 3 & 1 & 24 \\ 
\enddata 
\tablecomments{Source Type gives the spectroscopic classification of 
  the sources.} 
\end{deluxetable} 
 
\section{Galaxy Sample} 
In Table \ref{xlfsampletab} we give the identification numbers (as 
given in Giacconi et  
al. 2002 and Alexander et al. 2003), redshift and $\log L_X$ values for the 
galaxies in the Bayesian-selected sample.  The sources that are also 
in the spectroscopic CDF-S sample are marked.  There are also 3 
additional sources in the spectroscopic CDF-S sample that were not 
selected as galaxies by the Bayesian analysis, which have IDs 49, 242, 
and 519, redshifts 0.53, 1.03, and 1.03, and $\log L_X$ values 42.3, 
42.3, and 41.9.  As might be expected, these were the spectroscopic 
CDF-S sample galaxies with the highest $L_X$ values. 
\begin{deluxetable}{rrc} 
%\rotate 
\tablecolumns{3} 
\tabletypesize{\footnotesize} 
\tablewidth{0pt} 
\tablecaption {XLF galaxy sample \label{xlfsampletab}}  
% 
% \small 
% 
\tablehead{ 
\colhead{ID \#$^{\rm a}$} & 
\colhead{$z$} & 
\colhead{$L_{X}$ (0.5--2~keV)}  
} 
\startdata 
 
\multicolumn{3}{c}{CDF-S sample:} \\ 
\hline 
  1 & 0.347 & 41.73 \\ 
 12* & 0.251 & 41.65 \\ 
 40 & 0.545 & 42.16 \\ 
 77 & 0.622 & 42.26 \\ 
 84 & 0.103 & 40.38 \\ 
 95 & 0.076 & 39.95 \\ 
 96 & 0.274 & 40.78 \\ 
 97 & 0.180 & 41.16 \\ 
 98 & 0.279 & 41.02 \\ 
103 & 0.215 & 40.92 \\ 
116 & 0.076 & 39.87 \\ 
170 & 0.664 & 41.55 \\ 
173* & 0.524 & 40.92 \\ 
175* & 0.522 & 41.12 \\ 
185 & 0.925 & 41.51 \\ 
186 & 1.158 & 41.81 \\ 
211 & 0.679 & 41.81 \\ 
218 & 0.497 & 41.29 \\ 
224* & 0.738 & 41.77 \\ 
229* & 0.103 & 39.78 \\ 
233* & 0.577 & 41.01 \\ 
236 & 0.731 & 41.68 \\ 
246 & 0.690 & 41.98 \\ 
247 & 0.040 & 38.64 \\ 
249 & 0.964 & 41.85 \\ 
504 & 0.541 & 41.53 \\ 
509 & 0.556 & 41.59 \\ 
511 & 0.767 & 41.52 \\ 
512* & 0.668 & 41.52 \\ 
514 & 0.103 & 39.63 \\ 
516* & 0.665 & 41.47 \\ 
521* & 0.131 & 39.94 \\ 
525 & 0.229 & 40.49 \\ 
534 & 0.676 & 41.03 \\ 
535* & 0.575 & 41.43 \\ 
536 & 0.444 & 41.10 \\ 
538 & 0.310 & 40.50 \\ 
552 & 0.673 & 41.63 \\ 
553 & 0.366 & 40.97 \\ 
554 & 0.225 & 40.67 \\ 
556 & 0.635 & 41.33 \\ 
557 & 0.500 & 40.85 \\ 
558 & 0.585 & 41.49 \\ 
559 & 0.114 & 39.78 \\ 
560* & 0.669 & 41.36 \\ 
565* & 0.363 & 40.42 \\ 
566 & 0.734 & 41.74 \\ 
567* & 0.456 & 40.81 \\ 
573* & 0.414 & 40.61 \\ 
575* & 0.340 & 40.45 \\ 
577* & 0.547 & 41.23 \\ 
578 & 0.969 & 41.62 \\ 
580* & 0.664 & 41.13 \\ 
581 & 0.799 & 41.44 \\ 
582* & 0.241 & 40.15 \\ 
586* & 0.580 & 41.05 \\ 
587* & 0.246 & 40.07 \\ 
590 & 0.280 & 40.36 \\ 
592 & 1.150 & 41.78 \\ 
594 & 0.733 & 41.91 \\ 
617 & 0.588 & 41.54 \\ 
619 & 0.050 & 38.91 \\ 
620* & 0.648 & 41.15 \\ 
621 & 0.290 & 40.30 \\ 
624* & 0.668 & 41.17 \\ 
625 & 1.151 & 41.76 \\ 
627* & 0.248 & 40.12 \\ 
628 & 0.200 & 39.98 \\ 
629 & 0.410 & 40.71 \\ 
644 & 0.119 & 40.08 \\ 
646 & 0.438 & 40.79 \\ 
650 & 0.223 & 40.67 \\ 
651 & 0.182 & 40.57 \\ 
652* & 0.077 & 39.24 \\ 
\enddata 
\tablenotetext{a}{ID number for CDF-S sample is the XID as listed in Giacconi et al. (2002).   
} 
\tablenotetext{*}{These sources are also in the spectroscopic CDF-S sample.}
\end{deluxetable} 
%\hline

\begin{deluxetable}{rrcrrc} 
%\rotate 
\tablecolumns{6} 
\tabletypesize{\footnotesize} 
\tablewidth{0pt} 
\tablecaption {XLF galaxy sample}  
\tablehead{ 
\colhead{ID \#$^{\rm a}$} & 
\colhead{$z$} & 
\colhead{$L_{X}$ (0.5--2~keV)} &
\colhead{ID \#$^{\rm a}$} & 
\colhead{$z$} & 
\colhead{$L_{X}$ (0.5--2~keV)}  
} 
\startdata 

\multicolumn{6}{c}{CDF-N sample:} \\ 
\hline 
  3 & 0.138 & 40.29 &
  6 & 0.135 & 40.83 \\ 
 22 & 0.317 & 39.89 & 
 46 & 0.207 & 40.43 \\ 
 49 & 0.296 & 41.27 & 
 55 & 0.637 & 41.30 \\ 
 56 & 0.108 & 39.77 & 
 57 & 0.375 & 40.87 \\ 
 60 & 0.529 & 40.78 & 
 62 & 0.087 & 38.70 \\ 
 66 & 0.333 & 41.24 & 
 67 & 0.639 & 41.39 \\ 
 69 & 0.520 & 41.11 & 
 78 & 0.747 & 41.66 \\ 
 81 & 0.409 & 40.77 & 
 87 & 0.136 & 39.76 \\ 
 93 & 0.275 & 42.08 & 
101 & 0.454 & 40.81 \\ 
103 & 0.969 & 42.13 & 
105 & 0.319 & 39.99 \\ 
111 & 0.515 & 40.64 & 
114 & 0.534 & 40.63 \\ 
119 & 0.473 & 40.83 & 
120 & 0.694 & 41.27 \\ 
121 & 0.520 & 41.33 & 
126 & 0.779 & 40.91 \\ 
131 & 0.631 & 40.62 & 
132 & 0.647 & 41.07 \\ 
136 & 0.472 & 40.65 & 
138 & 0.483 & 40.79 \\ 
141 & 0.746 & 41.87 & 
148 & 1.130 & 41.96 \\ 
166 & 0.455 & 40.50 & 
169 & 0.845 & 41.53 \\ 
175 & 1.014 & 41.99 & 
177 & 1.016 & 42.14 \\ 
180 & 0.456 & 41.24 & 
189 & 0.410 & 41.01 \\ 
192 & 0.680 & 40.80 & 
197 & 0.081 & 38.92 \\ 
200 & 0.971 & 41.31 & 
203 & 1.143 & 41.27 \\ 
207 & 0.300 & 41.04 & 
209 & 0.510 & 41.36 \\ 
210 & 0.848 & 41.50 & 
211 & 0.846 & 41.35 \\ 
212 & 0.943 & 41.77 & 
214 & 1.144 & 41.37 \\ 
215 & 1.006 & 41.17 & 
218 & 0.089 & 39.31 \\ 
219 & 0.845 & 40.96 & 
224 & 0.700 & 41.92 \\ 
225 & 0.290 & 40.30 & 
227 & 0.556 & 40.64 \\ 
230 & 1.011 & 41.62 & 
234 & 0.454 & 40.85 \\ 
241 & 0.851 & 41.83 & 
244 & 0.971 & 41.36 \\ 
245 & 0.321 & 39.90 & 
249 & 0.475 & 41.30 \\ 
251 & 0.139 & 39.59 & 
255 & 0.113 & 40.31 \\ 
256 & 0.593 & 41.09 & 
257 & 0.089 & 38.92 \\ 
258 & 0.752 & 40.96 & 
260 & 0.475 & 40.72 \\ 
264 & 0.319 & 40.26 & 
265 & 0.410 & 40.74 \\ 
269 & 0.358 & 40.36 & 
274 & 0.321 & 40.92 \\ 
279 & 0.890 & 41.07 & 
280 & 0.850 & 41.12 \\ 
282 & 0.202 & 39.87 & 
285 & 0.288 & 41.03 \\ 
288 & 0.792 & 41.49 & 
291 & 0.517 & 40.59 \\ 
294 & 0.474 & 41.42 & 
295 & 0.845 & 41.31 \\ 
296 & 0.663 & 41.11 & 
300 & 0.137 & 39.54 \\ 
305 & 0.299 & 40.15 & 
308 & 0.515 & 40.53 \\ 
310 & 0.761 & 40.91 & 
311 & 0.914 & 41.33 \\ 
313 & 0.800 & 41.30 & 
316 & 0.231 & 40.32 \\ 
320 & 0.956 & 41.38 & 
326 & 0.359 & 40.30 \\ 
327 & 0.913 & 41.16 & 
332 & 0.561 & 40.70 \\ 
333 & 0.377 & 41.62 & 
337 & 0.902 & 41.04 \\ 
339 & 0.253 & 39.95 & 
346 & 1.017 & 40.98 \\ 
351 & 0.940 & 41.59 & 
353 & 0.422 & 40.80 \\ 
354 & 0.568 & 41.05 & 
355 & 1.027 & 41.77 \\ 
356 & 0.956 & 42.20 & 
358 & 0.907 & 41.86 \\ 
359 & 0.902 & 41.62 & 
378 & 1.084 & 41.40 \\ 
383 & 0.105 & 39.18 & 
387 & 1.081 & 41.70 \\ 
388 & 0.559 & 41.05 & 
389 & 0.557 & 40.78 \\ 
392 & 0.411 & 40.51 & 
395 & 0.411 & 40.77 \\ 
401 & 0.935 & 41.51 & 
404 & 0.104 & 39.26 \\ 
407 & 1.200 & 41.94 & 
410 & 0.113 & 39.30 \\ 
414 & 0.800 & 41.46 & 
415 & 0.116 & 39.60 \\ 
418 & 0.279 & 40.35 & 
425 & 0.214 & 40.11 \\ 
426 & 0.159 & 39.75 & 
428 & 0.298 & 40.47 \\ 
433 & 1.022 & 41.64 & 
435 & 0.201 & 39.99 \\ 
436 & 0.189 & 40.38 & 
438 & 0.220 & 40.46 \\ 
443 & 0.231 & 40.69 & 
446 & 0.410 & 40.56 \\ 
450 & 0.935 & 41.74 & 
453 & 0.838 & 42.11 \\ 
454 & 0.458 & 41.48 & 
458 & 0.069 & 39.55 \\ 
460 & 1.084 & 42.17 & 
462 & 0.511 & 40.44 \\ 
466 & 0.440 & 40.81 & 
469 & 0.188 & 40.09 \\ 
471 & 1.170 & 41.90 & 
476 & 0.475 & 40.95 \\ 
480 & 0.456 & 41.19 & 
489 & 1.024 & 42.24  
\enddata 
\tablenotetext{a}{   
For the CDF-N sample, it is the ID number listed in Alexander et al. (2003). } 
\end{deluxetable}

\end{document}